**Acute Respiratory Distress Syndrome is a TH17-like and Treg immune disease**

Wan-Chung Hu*
*Postdoctorate, Genomics Research Center, Academia Sinica, , Taipei, Taiwan
Department of Neurology, Shin Kong Memorial Hospital, Taipei, Taiwan

## Abstract

Acute Respiratory Distress Syndrome (ARDS) is a very severe syndrome leading to respiratory failure and subsequent mortality. Sepsis is one of the leading causes of ARDS. Thus, extracellular bacteria play an important role in the pathophysiology of ARDS. Overactivated neutrophils are the major effector cells in ARDS. Thus, extracellular bacteria triggered TH17-like innate immunity with neutrophil activation might accounts for the etiology of ARDS. Here, microarray analysis was employed to describe TH17-like innate immunity-related cytokine including TGF-β and IL-6 up-regulation in whole blood of ARDS patients. It was found that the innate TH17-related TLR1,2,4,5,8, HSP70, G-CSF, GM-CSF, complements, defensin, PMN chemokines, cathepsins, Fc receptors, NCFs, FOS, JunB, CEBPs, NFkB, and leukotriene B4 are all up-regulated. TGF-β secreting Treg cells play important roles in lung fibrosis. Up-regulation of Treg associated STAT5B and TGF-β with down-regulation of MHC genes, TCR genes, and co-stimulation molecule CD86 are noted. Key TH17 transcription factors, STAT3 and RORα, are down-regulated. Thus, the full adaptive TH17 helper CD4 T cells may not be successfully triggered. Many fibrosis promoting genes are also up-regulated including MMP8, MMP9, FGF13, TIMP1, TIMP2, PLOD1, P4HB, P4HA1, PDGFC, HMMR, HS2ST1, CHSY1, and CSGALNACT. Failure to induce successful adaptive immunity could also attribute to ARDS pathogenesis. Thus, ARDS is actually a TH17-like and Treg immune disorder.

## Introduction

Acute respiratory distress syndrome (ARDS) is a severe cause of respiratory failure. Despite of current treatment, the mortality rate remains very high. We still don't have successful management strategies to deal with ARDS. Most important of all, we still don't know the exact pathophysiology of ARDS. Sepsis or bacteremia is the leading cause of ARDS. Besides, neutrophil activation is reported in many studies in the lungs of ARDS patients. It was hypothesized that extracellular bacteria-induced TH17 immunity overactivation might be the etiology of ARDS. Microarray analysis to will be used to study the immune-related gene profiles in peripheral leukocytes of ARDS patients.

According to Harrison's internal medicine, the time course of ARDS can be divided into three stages.(1) First, the exudative phase. In this phase, injured alveolar capillary endothelium and type I pneumocytes cause the loss of tight alveolar barrier.

Thus, edema fluid rich in protein accumulate in the interstitial alveolar spaces. It has been reported that cytokines (IL-1, IL-6, and TNF-α) and chemokines (IL-8, and leukotriene B4) are present in lung in this phase (2, 3). A great numbers of neutrophils traffic into the pulmonary interstitium and alveoli.(4, 5) Alveolar edema predominantly leads to diminished aeration and atelectasis. Hyaline membranes start to develop. Then, intrapulmonary shunting and hypoxemia develop. The situation is even worse with microvascular occlusion which leads to increasing dead space and pulmonary hypertension. The exudative phase encompasses the first seven days of disease after exposure to a precipitating ARDS risk factor such as sepsis, aspiration pneumonia, bacterial pneumonia, pulmonary contusion, near drowning, toxic inhalation injury, severe trauma, burns, multiple transfusions, drug overdose, pancreatitis, and post-cardiopulmonary bypass.

Second, proliferative phase. This phase usually lasts from day 7 to day 21. Although many patients could recover during this stage, some patients develop progressive lung injury and early change of pulmonary fibrosis. Histologically, this phase is the initiation of lung repair, organization of alveolar exudates, and a shift from a neutrophil to a lymphocyte dominant pulmonary infiltrate. There is a proliferation of type II pneumocytes which can synthesize new pulmonary surfactants. They can also differentiate into type I pneumocytes. In addition, there is beginning of type III procollagen peptide presence which is the marker of pulmonary fibrosis.

Third, fibrotic phase. Although many patients with ARDS recover lung function three weeks after the initial lung injury, some enter a fibrotic phase that may require long term support on mechanical ventilators. Histologically, the alveolar edema and inflammatory exudates in early phases are converted to extensive alveolar duct and interstitial fibrosis. Intimal fibroproliferation in the pulmonary microvascular system leads to progressive vascular occlusion and pulmonary hypertension.

**Material and Methods**

Microarray dataset

According to Dr. J. A. Howrylak's research in Physiol Genomics 2009, who collected total RNA from whole blood in sepsis and sepsis-induced ARDS patients.(6) He tried to find out molecular signature of ARDS compared to sepsis patients. His dataset is available in Gene Expression Omnibus (GEO) www.ncbi.nlm.nih.gov/geo (accession number GSE 10474). The total number of his sepsis-induced ARDS is 13. The overall mortality of these patients is 38%. The second dataset is from GSE20189 of Gene Expression Omnibus. This dataset was collected by Dr. Melissa Rotunno in Cancer Prevention Research 2011.(7) Molecular signature of early stage of lung adenocarcinoma was studied by microarray. We use the healthy control (sample size 21) whole blood RNA from this dataset to compare the ARDS patients. In this study, we perform further analysis to study peripheral leukocyte gene expression profiles of ARDS compared to those of healthy controls.

Statistical analysis

Affymetrix HG-U133A 2.0 genechip was used in both samples. RMA express software (UC Berkeley, Board Institute) was used to do normalization and to rule out the outliners of the above dataset. The potential outliners of samples were removed according to the following criteria:

1. samples which have strong deviation in NUSE plot
2. samples which have broad spectrum in RLE value plot
3. samples which have strong deviation in RLE-NUSE mutiplot
4. samples which exceed 99% line in RLE-NUSE T2 plot.

Then, Genespring XI software was done to analyze the significantly expressed genes between ARDS and healthy control leukocytes. P value cut-off point is set to be less than 0.05. Fold change cut-off point is >2.0 fold change. Benjamini-hochberg corrected false discovery rate was used during the analysis. Totally, a genelist of 3348 genes was generated from the HGU133A2.0 chip with 18400 transcripts including 14500 well-characterized human genes.

RT-PCR confirmation

Dr. J. A. Howrylak performed real time PCR for selected transcripts (cip1, kip2) by using TaqMan Gene Expression Assays (Applied Biosystems, Foster City, CA). In the second dataset, Dr. Melissa Rotunno also performed qRT-PCR test to validate the microarray results. RNA quantity and quality was determined by using RNA 600 LabChip-Aligent 2100 Bioanalyzer. RNA purification was done by the reagents from Qiagen Inc. All real-time PCRs were conducted by using an ABI Prism 7000 Sequence Detection System with the designed primers and probes for target genes and an internal control gene-GAPDH. This confirms that their microarray results are convincing compared to RT-PCR results.

**Results**

RMA analysis of whole blood from healthy normal control

The RMA analysis was performed for RNA samples from whole blood of healthy control of the lung adenocarcinoma dataset. Raw boxplot, NUSE plot, RLE value plot, RLE-NUSE multiplot, and RLE-NUSE T2 plot were generated. Then, sample was included and excluded by using these graphs (Figure 1A, 1B, 1C, 1D, 1E). Because of the strong deviation in the T2 plot, the sample GSM506435 was removed for the further analysis.

RMA analysis of whole blood from acute lung injury patients

The RMA analysis was performed for RNA samples from whole blood of healthy control of the ARDS dataset. Raw boxplot, NUSE plot, RLE value plot, RLE-NUSE multiplot, and RLE-NUSE T2 plot were generated. Then, sample was included and excluded by using these graphs (Figure 2A, 2B, 2C, 2D, 2E)

TH17-like innate immunity and Treg-related genes are up-regulated in ARDS

Based on the microarray analysis, we found out that many TH17-related genes are up-regulated in ARDS including Toll-like receptors 1,2,4,5,8, complement, heat shock protein 70, cathpesin, S100A proteins, leukotrienes, defensins, TH17-related chemokines, and MMPs. Many fibrosis related genes are also up-regulated including key collagen synthesis enzymes and fibroblast growth factor. Key TH17 initiating cytokines including TGF beta and IL-6 are also up-regulated. This explains that TH17-like immunity is initiated in ARDS. NK cell and T cell related genes are down-regulated. This explains that TH1 or THαβ immunological pathway is not triggered in ARDS. In addition, TGFB1 and STAT5B up-regulations suggest that Treg cells play important roles in ARDS pathogenesis. (Table 1-14)

In Table 1, it can be seen that the TLR1, 2,4,5,8 were up-regulated with the expression of IRAK4. Thus, toll-like receptor signaling is generated during ARDS. It is worth noting that TLR1,2,4,5,8 are all anti-bacteria TH17 innate immunity signaling (TLR8 is against CpG rich oligonucleotides). Thus, strong TH17-like related Toll signaling can be activated.

In Table 2, differentiated expression of heat shock proteins can be seen. Most important of all, HSPA1A, HSPA1B, and HSPA4 are up-regulated. HSPA1A and HSPA1B are greater than 6.0 fold up-regulation. These heat shock proteins are HSP70 family which can activate TH17 related toll-like receptors (TLR2 and TLR4) to generate anti-bacterial immunity. Because of the co-upregulation of TLR2/4 and HSP70, it is evident that the proinflammatory signaling is generated in acute lung injury.

In Table 3, chemokine and chemokine receptor genes are differentially regulated. TH1 T cell chemotaxis factor CCL5 with its receptor CCR1 and CCL4 with its receptor CCR5 are both down-regulated. TH2 eosinophil chemotaxic factor receptor CCR3 and THαβ NK cell chemotaxic factors XCL1 and XCL2 are all down-regulated. However, TH17-related chemokines such as PAF related molecules and S100A binding proteins are up-regulated. These findings suggest that TH17-like related response to recruit neutrophils is initiated. Toll-like receptor signaling can activate these TH17-like chemokines.

In Table 4, strikingly and surprisingly, all the MHC related genes are down-regulated during acute lung injury. These genes include HLA-DRB, HLA-DRA, HLA-DQB, HLA-DPA, HLA-DQA, and HLA-DMB. The HLA-DQA1 has the lowest expression level with greater than 5.8 fold down-regulation. Thus, MHC antigen presentation genes are down-regulated in acute lung injury.

In Table 5, many immune-related transcription factors are up-regulated or down-regulated. Key Treg related key transcription factor, STAT5B, is up-regulated(8). And, THαβ and TH1 related key transcription factor, STAT1, is down-regulated. In addition, TH2 related transcription factor, GATA3, is also down-regulated. Surprisingly, key TH17 transcription factors, STAT3 and RORα, are down-regulated.(9) These findings suggest that full TH17 immunity may not be activated during ARDS. Other innate immunity related genes for myeloid or granulocyte lineages are up-regulated including AP1 (Fos and Jun), CEBP family genes, and NFIL3. It is worth noting that the inhibitor of NFkB, key innate immunity mediator, is down-regulated in acute lung injury. Besides, T cell related transcription factors including NFATC3, NFAT5, and NFATC2IP are down-regulated in ARDS. JAK2, a signal transduction for all cytokines and STATs proteins, is also up-regulated. STAT5B is the key transcription factor for regulatory T cells. And, down-regulation of RORα and STAT3 means only TH17-like innate immunity is activated in acute lung injury.

In Table 6, it can be seen that many chemotaxic factors, leukotrienes and prostaglandins, are up-regulated or down-regulated. The key enzyme: leukotriene A4 hydrolase for leukotriene B4, a potent PMN chemoattractant, is up-regulated. Besides, leukotriene B4 receptor is also up-regulated. Furthermore, the receptor of PGD2, a TH2 related effector molecule, is 9 fold down-regulated. In addition, the gene 15-hydroxyprostaglandin dehydrogenase (HPGD), which is responsible for shutting down prostaglandin, is 27 fold up-regulated. Key molecules including phospholipase A 2 and arachidonate 5-lipoxygenase to initiate leukotriene synthesis are also up-regulated in ARDS. And, formyl peptide receptor 2, another chemotaxic receptor of PMN, is also up-regulated. Innate immune initiators, SERPINB1 and SERPINB2, are also up-regulated in acute lung injury.

In Table 7, many fibrosis related genes are up-regulated during ARDS. Most strikingly, Matrix metalloproteinase 8 (MMP8) has greater than 28 fold up-regulation. Matrix metalloproteinase 9 (MMP9) has greater than 11 fold up-regulation. MMP8 & MMP9 play key roles in the pathogenesis of ARDS(10). In addition, MMP25, TIMP1, and TIMP2 are also up-regulated in ARDS. Thus, severe extracellular matrix destruction happens in ARDS. Fibroblast growth factors including FGF13 (5 fold up-regulation) and PDGFc (12 fold up-regulation) are also significantly expressed. Chondroitin sulfate deposition is reported in pulmonary fibrosis (11). In this study, chondroitin sulfate synthetase and chondroitin sulfate N-acetylgalactosaminyltransferase 1 & 2 are also found up-regulated. There is 7.5 fold up-regulation in chondroitin sulfate N-acetylgalactosaminyltransferase 1 and 13 fold up-regulation in chondroitin sulfate N-acetylgalactosaminyltransferase 2. Carboxypeptidase D, which can up-regulate nitric oxide, is also up-regulated in acute lung injury.(12) Previous studies also found up-regulation of iNOS as well as nitric oxide during the inflammation in ARDS.(13) Heparanse and heparan sulfate 2-O-sulfotransferase 1 are also up-regulated in ARDS. Heparan sulfate sulfation is a potent stimulation of FGF signaling to cause fibrosis (14, 15). Hyaluronan-mediated motility receptor (RHAMM) is also up-regulated. Hyaluronan plays an important role in pulmonary fibrosis. In addition, key collagen synthesis enzymes, prolyl-4-hydroxylases and procollagen-lysine 1, 2-oxoglutarate 5-dioxygenase are also up-regulated in ARDS.(16, 17)

In Table 8, many complement related genes are up-regulated including CD59, C1QB, ITGAM, CR1, C3AR1, ITGAX, C1QA, C1RL, and C5AR1. Complements are important in the TH17 arm to kill extracellular bacteria. Thus, the whole complement machinery is activated in ARDS. In addition, two defensin genes, DEFA1B3 and DEFA4, are also up-regulated to defend the possible bacterial infection. Neutrophil overactivity with up-regulated complements and defensins plays a vital role in acute lung injury.

In Table 9, many cathepsin genes are up-regulated in acute respiratory distress syndrome including CTSK, CTSG, CTSZ, CTSA, CTSD, and CTSC except CTSO and CTSW. Cathepsins are important proteases in antigen processing. Thus, antigen processing is likely to be activated in ARDS. However, MHC related genes are down-regulated in acute lung injury. In addition, myeloperoxidase, which is the enzyme responsible for ingested bacteria killing, is upregulated in ARDS. Neutrophil cytosolic factor 1 & 4, the subunit of NADPH oxidase for ingested bacteria killing, are also up-regulated in ARDS.

In Table 10, two CSF receptors are up-regulated in acute lung injury. CSF2 receptor (GM-CSF receptor) is more than two fold up-regulated. And, CSF3 receptor (G-CSF receptor) is also more than two fold up-regulated. GM-CSF and G-CSF can stimulate granulocyte and monocyte proliferation. It means that myeloid and granulocyte lineages are proliferative in ARDS.

In Table 11, Fc receptor related genes including IgG Fc receptor 2A, IgA Fc receptor, IgG Fc receptor 2C, IgG Fc receptor 1B, and IgG Fc receptor 1A/1C are up-regulated in ARDS. These Fc receptors (CD32 & CD64), expressed on myeloid lineages, are related to TH17-related immunological pathway. Besides, TH2 related IgE Fc receptor 1A is down-regulated. These findings support that TH17-like armed innate immune response is activated in acute lung injury.

In Table 12, many TH17-like related cytokine genes are up-regulated in ARDS. Most importantly, the central TH17 cytokine initiators, TGFB1 and IL-6, are up-regulated in ARDS. Thrombospondin (THBS1), the activator of TGFB, is also strongly up-regulated. In addition, TH22 related cytokines such as IL32 and IL1A are down-regulated. And, IL1RN, an IL1 antagonist, is up-regulated. The receptors of TH17 immunity are also down-regulated, including IL17RA, IL6R, and TGFBR3. The receptors for Treg pathway are also down-regulated, including IL2RB and TGFBR3. However, the key TH17 downstream IL-6/TGFβ inducing transcription factors, STAT3 and RORα, are down-regulated. These findings suggest that only TH17-like innate immunity is up-regulated. Other immunological pathway cytokine receptors are up-regulated, including IL1RA, IL1R2, IL1R1, IL4R, IL18R1, IFNGR1, IFNGR2, IFNAR1, and TNFRSF1A. Besides, TH1 or THαβ associated interferon related genes are down-regulated, including ISG20L2, IFI16, GVINP1, GBP1, IFI44L, and IFIT3. In addition, FAS is up-regulated and Fas apoptotic inhibitory molecule 3 (FAIM3) is down-regulated. These findings suggest that apoptosis machinery is activated in ARDS.

In Table 13, many important CD molecules are differentially regulated. CD molecules are important in mediating host immune reaction. Thus, the up-regulation or down-regulation of these CD molecules suggests the status of host immunity. From this table, we can see that many T cell activation molecules are down-regulated, including CD8A, CD3G, CD3D, and CD86. CD8A is the molecule of cytotoxic T cell activation. CD3G is the molecule for helper or killer T cell activation. CD86 is the key co-stimulation signal to activate B cells and T cells. Thus, the adoptive immunity including B and T lymphocytes are not likely to be activated in ARDS.

In the Table 14, many NK cells and T cell related molecules are seen to be down-regulated. Molecules related to NK cell activation include grazymes (GZMK, GZMM, GZMB, and GZMH), perforin(PRF1), and killer cell receptors (KLRK1, KLRD1, KLRG1, KLRB1, NKTR, and KLRF1). NKTR is 13-fold down-regulated and granulysin (GNLY) is 5-fold down-regulated. NK cells are the key effector cells in THαβ immunity. Thus, THαβ immunity is not activated or even down-regulated in ARDS. In addition, many TCR related genes are also down-regulated, including TRBC1, TRAC/J17/V20, TRBC2, TRD@, TRAP/TRGC2, and TRDV3. Several TCR related genes have greater than 5-fold down-regulation. We can see the down-regulations of MHC genes, CD costimulation molecules, STAT3, and TCR genes as well as the up-regulation of TGFB1 and STAT5B. These findings suggest that T cells are not activated in the acute lung injury. Thus, adaptive immunity cannot be successfully triggered in ARDS. Because successful adaptive TH17 helper cells need at least three signals: TGFβ, IL-6, and TCR. The absence of TCR signal cannot fully trigger TH17 CD4 T cells. Our findings suggest that Treg plays an important role in ARDS pathophysiology.

**Discussion**

ARDS is a very severe respiratory complication. Sepsis is the major risk factor of ARDS. Sepsis is caused by the uncontrolled bacteremia due to bacterial infection. In addition, PMNs overactivation is very important in the pathogenesis of ARDS. Thus, extracellular bacteria induced TH17 immunity with neutrophil activation might be the key in the pathophysiology of ARDS.

Here, I propose a detail pathogenesis to explain the three stages of ARDS. In the first exudative stage, neutrophils are attracted to lung due to chemotaxic agents such as IL-8 or C5 or leukotriene B4. During sepsis, bacterial infection in pulmonary tissue can trigger pulmonary epithelial cells, pulmonary endothelial cells, pulmonary fibroblast, and alveolar macrophage to be activated. Toll-like receptors 1,2,4,5 as well as heat shock proteins (HSP60, HSP70) are key molecules to trigger TH17 host immunity (18-26). Heat shock proteins are important stress proteins in situation such as burn, trauma, hemorrhagic shock, near drowning or acute pancreatitis.(27-29) HSP60 and HSP70 can activate TH-17 related Toll-like receptors. Thus, TH17 related cytokines such as IL-17, IL-1, TNF-α, and IL-6 as well as TH-17 related chemokines such as IL-8 and other CXCL group chemokines will be triggered. Bacterial infection can let pulmonary epithelial cells to release chemokines and cytokines.(30-32) TH17 cytokines will start to activate TH17 immunity including activating PMN effector function for immunity against extracellular bacteria. The cytokine storm during ARDS is now explained. However, in this study, it was found that CD86 costimulation signal, TCR genes, and majority of MHC genes are down-regulated in ARDS. Thus, adaptive full TH17 immunity related effective and specific antibody and TCR response against bacteria may not be triggered. However, one research found that there is IL-8 autoantibody in ARDS patients.(33) IL-8 as well as leukotriene B4 is the main chemoattractant in pulmonary tissue.(34-40) It was first identified in lung giant cell lines.(41) Besides, IL-8 has high affinity to bind to the heparin sulfate and chondroitin sulfate-enriched lung tissue.(42, 43) And, IL-8 retention in pulmonary tissue can further recruit neutrophils to lung. This can explain why IL-8 secreted from distant site such as pancreas during acute pancreatitis can cause ARDS. However, IL-8 itself is not changed in this study. And, there are several limitations in the IL8 autoantibody study (33). First, anti-IL8-IL8 complex can be detected in 55% of healthy control serum. There is no significant difference of IL8-antiI-L8 complex between ARDS patients' serum and healthy controls' serum. In addition, IL8 autoantibody can suppress IL8 binding activity for neutrophils and it can reduce IL8's chemotaxic activity. Thus, IL8 autoantibody's importance in ARDS pathogenesis is doubtful. Besides, several studies can support the role of TH17 immunity in the pathogenesis of ARDS. G-CSF, the growth factor of neutrophils, can cause the common symptoms of ARDS.(44) And, suppression of NFkB can attenuate ARDS progression.(45, 46) Key THαβ cytokine, IL-10, can reduce the severity of ARDS.(47)

Bacterial infection is the most common risk factor of ARDS. However, certain pathogens other than bacteria also are risks for developing ARDS. Plasmodium falciparum malarial infection can also cause the complication of ARDS. The reason for this is that Plasmodium falciparum can activate heat shock proteins to trigger TH17 immunity to cause ARDS.(author's paper in press)(48) SARS-CoV and H1N1 Avian flu virus can also down-regulate normal anti-viral interferon-α/β and up-regulate TH17-like immunity to trigger ARDS. (author's paper in press: Viral Immunology)(49, 50) Thus, the above phenomena suggest that TH17 inflammation is the key to the pathogenesis of ARDS. If different pathogens lead to a common pathway of TH17-related innate immunity, they will cause the same consequence of ARDS. It is also seen in burn, trauma, or pancreatitis when TH17-like innate immunity is also activated.

In the second proliferative stage, lymphocytes replace neutrophils and become the dominant population in ARDS. These lymphocytes are TH17-like lymphocytes and Treg lymphocytes. TH17-related cytokines such as IL-17, IL-1, IL-6, and TNF-α can continue the inflammatory process due to the activation of innate immunity. However, once the bacterial antigen during sepsis is cleared, Toll-like receptor signaling is stopped, and no further proinflammatory cytokines such as IL-6 are synthesized. In addition, there is no TCR signal found in this study. Thus, TH17 adaptive immunity may not be successfully generated. This could be very important in ARDS pathogenesis. In TH17 immunity, both TGF-β and IL-6 are two important triggering cytokines. If there is no longer IL-6 signaling, only TGF-β is generated. IL-6 is the key factor to regulate the balance between Treg cells and TH17 cells. If there is enough IL-6, Treg cells will become TH17 cells. If there is not enough IL-6, TGF-β secreting Treg cells will be maintained. Treg cells are associated with STAT5B activation. Thus, in the third fibrosis stage, TGF-β secreting Treg cells are the dominant effector cells in ARDS.(51) In this study, patients' ARDS is induced by bacterial sepsis. Thus, failure to induce specific adaptive immunity such as TCR and antibody cannot successfully defeat these bacteria. In the early disease stage, overactive innate immunity with PMN activation causes severe lung consolidation. In the later stage, failure of adaptive immunity against bacteria causes the abundant regulatory T cells secreting TGF-β. TGF-β is a very strong fibrosis promoting agent and is the most important and potent stimulant in tissue fibrosis.(52, 53) TGF-β will promote the synthesis of multiple collagen genes.(54) Thus, overproduction of TGF-β in lung tissue will cause pulmonary fibrosis. TGF-β caused fibrosis is usually a process for repairing cavity after bacterial infection locus such as abscess. This mechanism can solve many controversial studies before. Several studies found that TLR4 and heat shock proteins can aggravate ARDS.(20) However, another studies found that TLR4 or heat shock protein can protect from pulmonary fibrosis after acute lung injury.(55-57) It is because TLR and heat shock signaling can maintain the activation of proinflammatory cytokines such as IL-6. Thus, no sole TGF-β overproduction happens in lung fibrosis. In an animal study, neutrophil inhibitor can attenuate the progression of acute lung injury (58). Thus, TH17 and Treg inflammatory process can fully explain the pathogenesis of ARDS.

Recently, two researches suggested that Treg cells play protective roles in acute lung

injury.(59, 60) We cannot agree with their suggestions. First of all, they used Rag-/- mice and found that ARDS is reduced in Rag-/- mice. However, both B cells and T cells are absent in Rag-/- mice. This finding can be explained that adaptive immune T & B lymphocytes relieve the TH17 innate ARDS pathological change. They also found that CD8 T cells have protective roles in ARDS. In this study, we found that antigen specific T cells are not activated in ARDS. Thus, specific TCR or antibody response against bacteria antigen could not be successfully triggered in acute lung injury. If the adaptive immunity can be triggered, it can limit ARDS pathogenesis. Second, they used a lung injury scoring system to assess the severity of ARDS. However, the scoring system only included lung congestion and inflammatory infiltration. The most important sequel of ARDS, pulmonary fibrosis, was not included in their studies. Thus, their conclusion that TGF beta secreting Treg cells can protect ARDS is questionable . Actually, anti-TGFB antibody can prevent mice from lung fibrosis.(61) After knowing the complete pathophysiology ofARDS, we may be able to develop better treatment strategies to managing this highly detrimental and fatal disease.

**Author's information**

Wan-Chung Hu is a MD from College of Medicine of National Taiwan University and a PhD from vaccine science track of Department of International Health of Johns Hopkins University School of Public Health. He is a postdoctorate in Genomics Research Center of Academia Sinica, Taiwan. His previous work on immunology and functional genomic studies were published at *Infection and Immunity* 2006, 74(10):5561, *Viral Immunology* 2012, 25(4):277, and *Malaria Journal* 2013,12:392. He proposed THαβ immune response as the host immune response against viruses.

Figure legends

Figure 1. RMA express plot for selecting samples in healthy controls.

1-A NUSE boxplot for normal control

1-B RLE boxplot for normal control

1-C RLE-NUSE multiplot for normal control

1-D RLE-NUSE T2 plot for normal control

1-E Raw data Boxpolt for normal control

Figure 2. RMA express plot for selecting samples in ARDS patients.

2-A NUSE boxplot for ARDS patients

2-B RLE boxplot for ARDS patients

2-C RLE-NUSE multiplot for ARDS patients

2-D RLE-NUSE T2 plot for ARDS patients

2-E Raw data Boxplot for ARDS patients

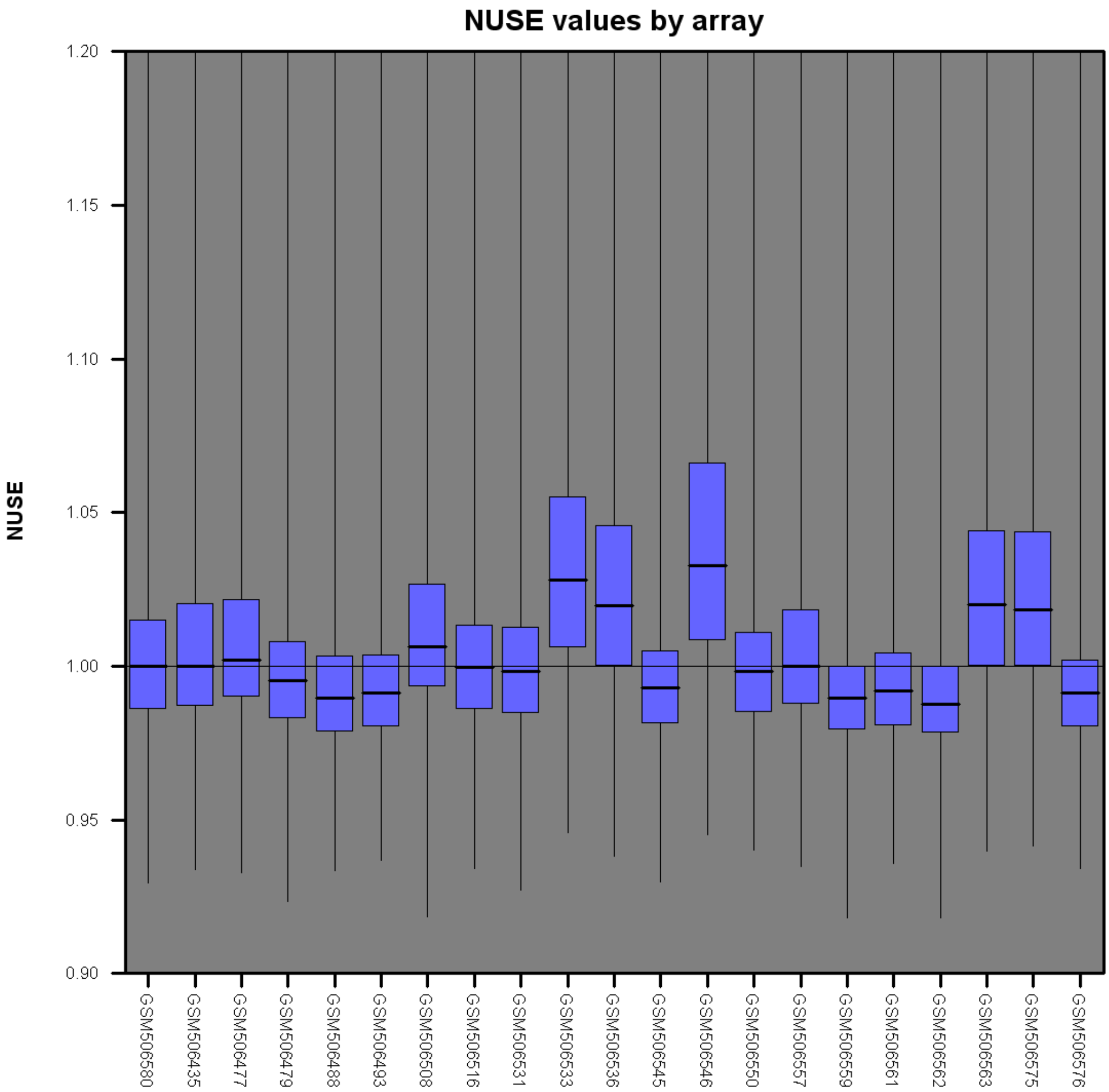

Figure 1-A

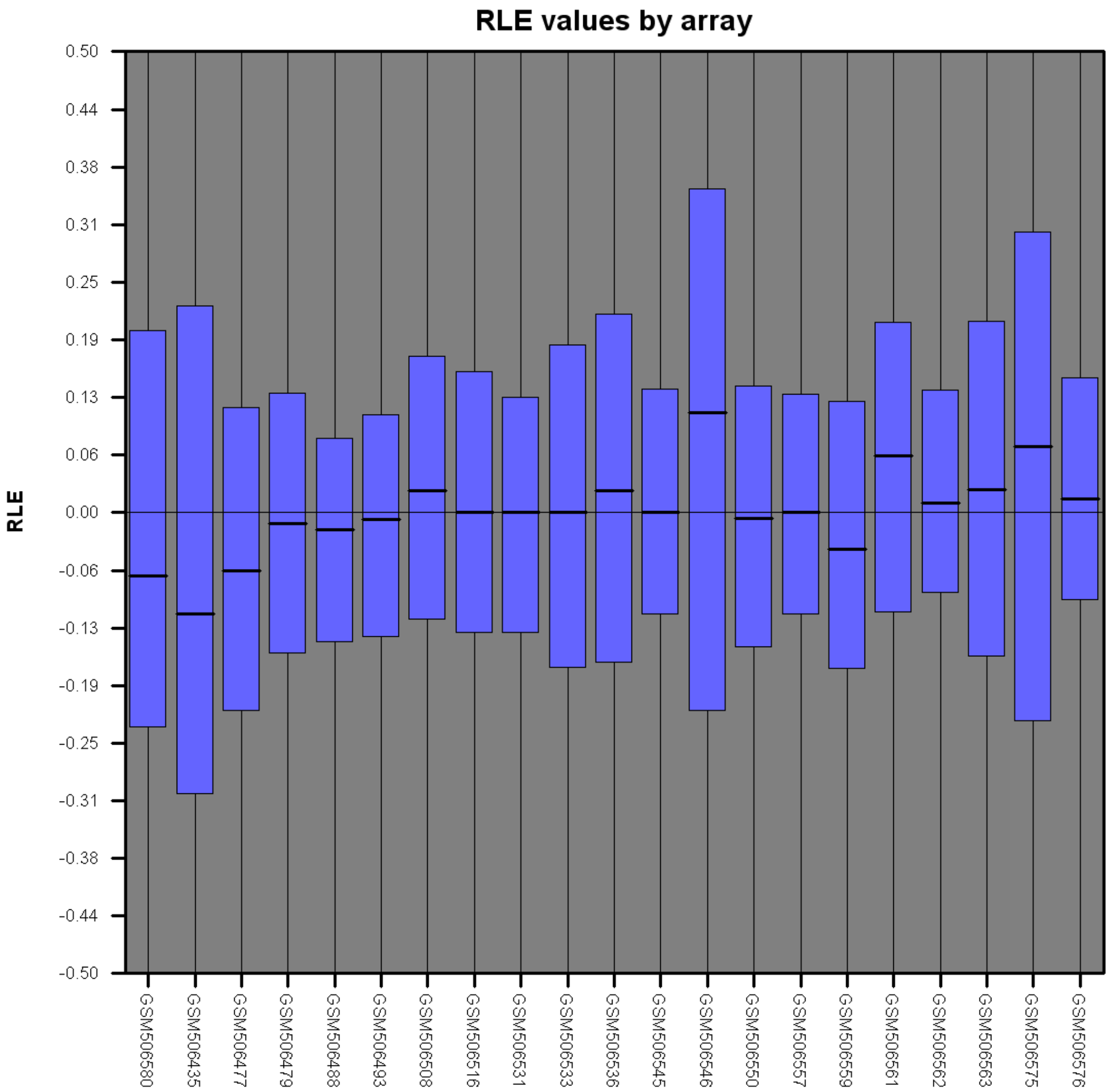

Figure 1-B

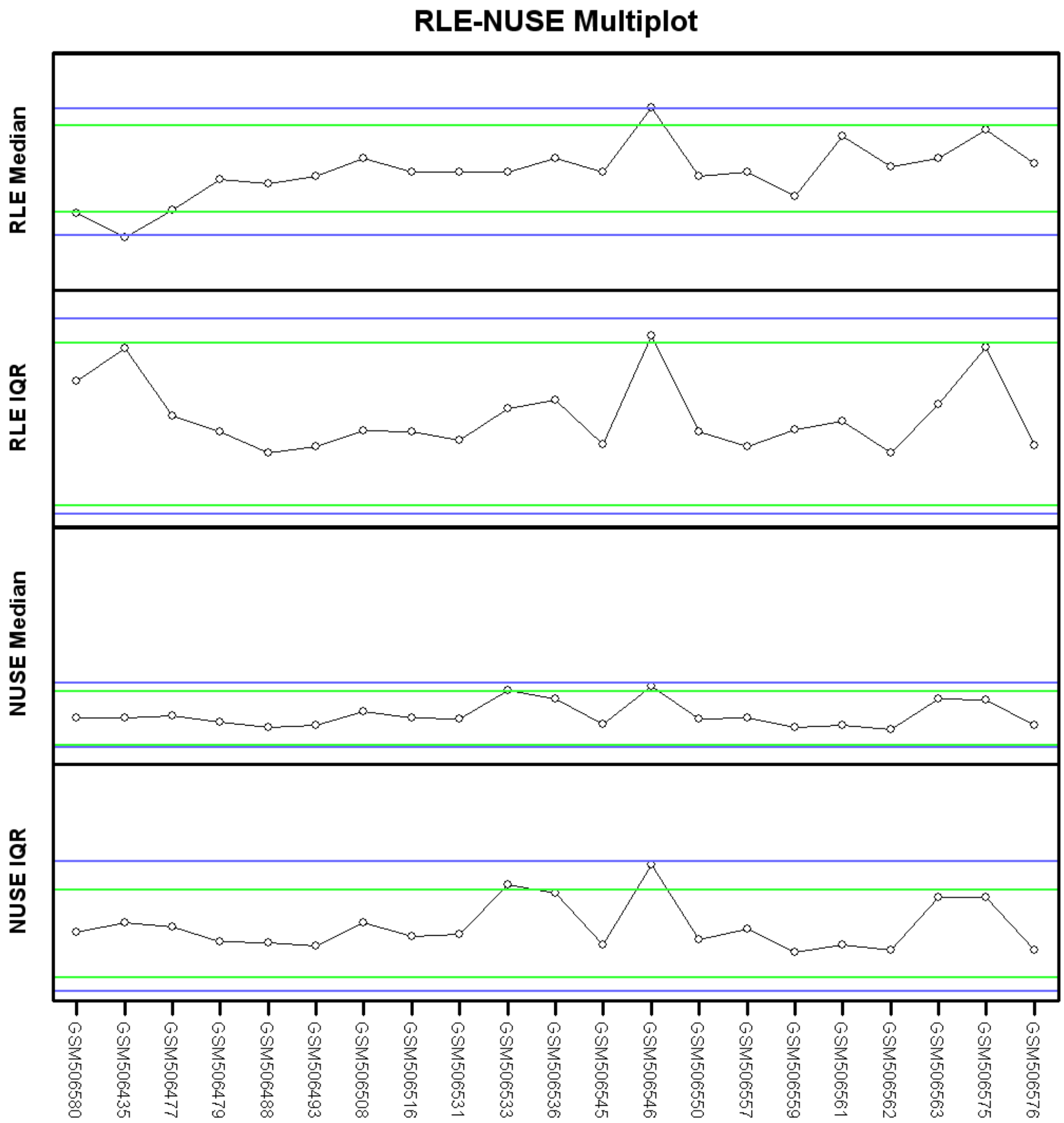

Figure 1-C

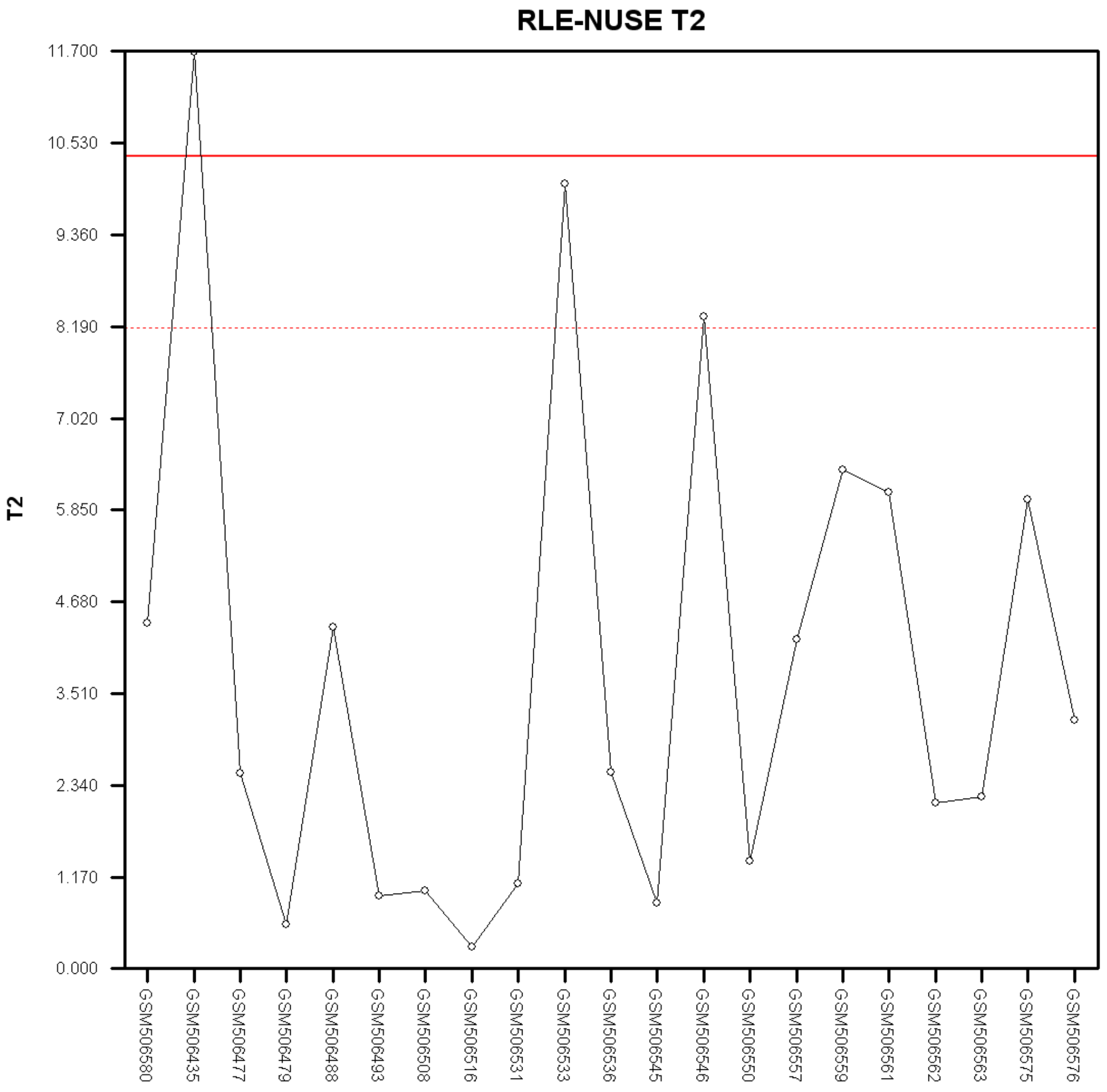

Figure 1-D

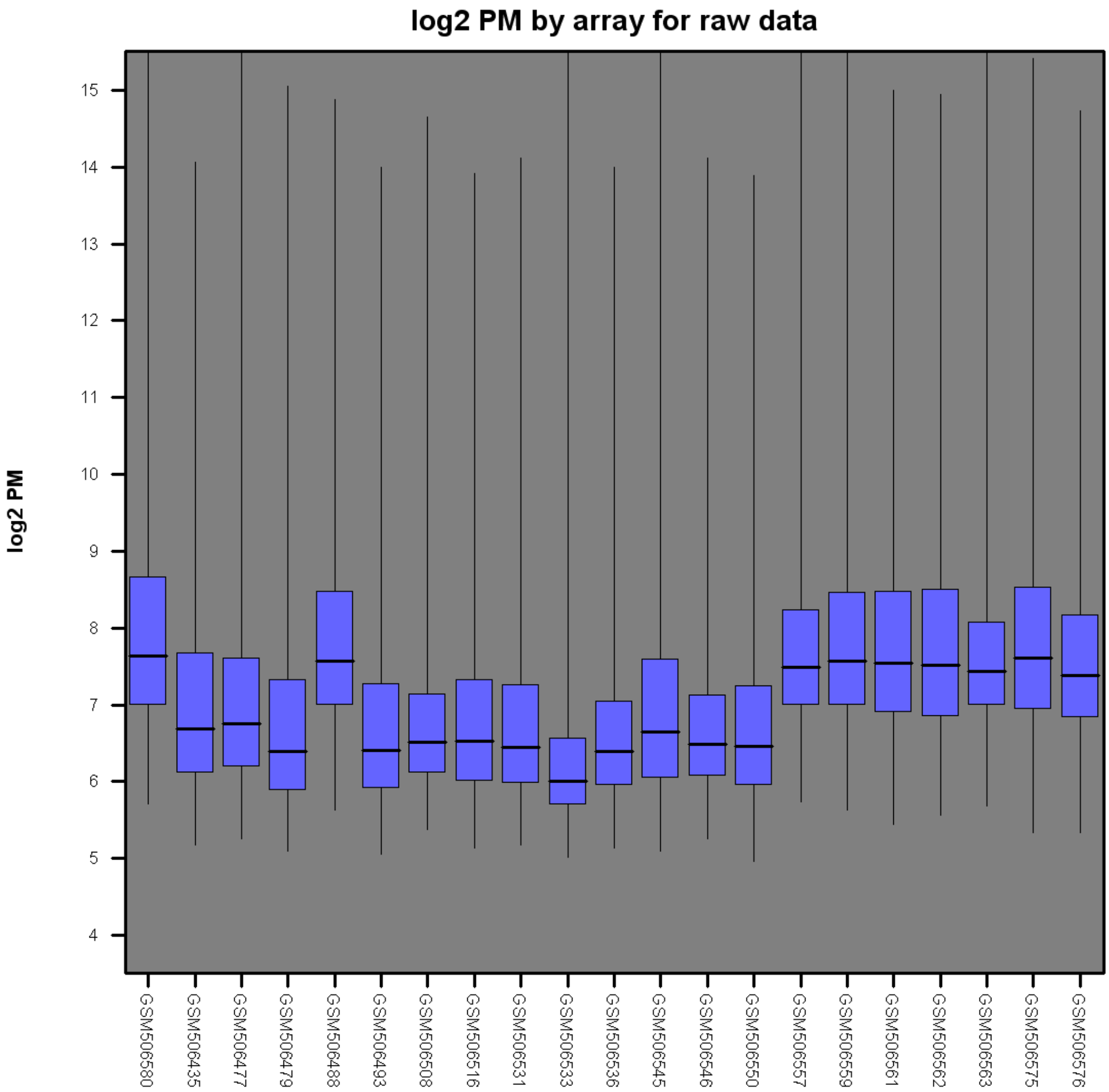

Figure 1-E

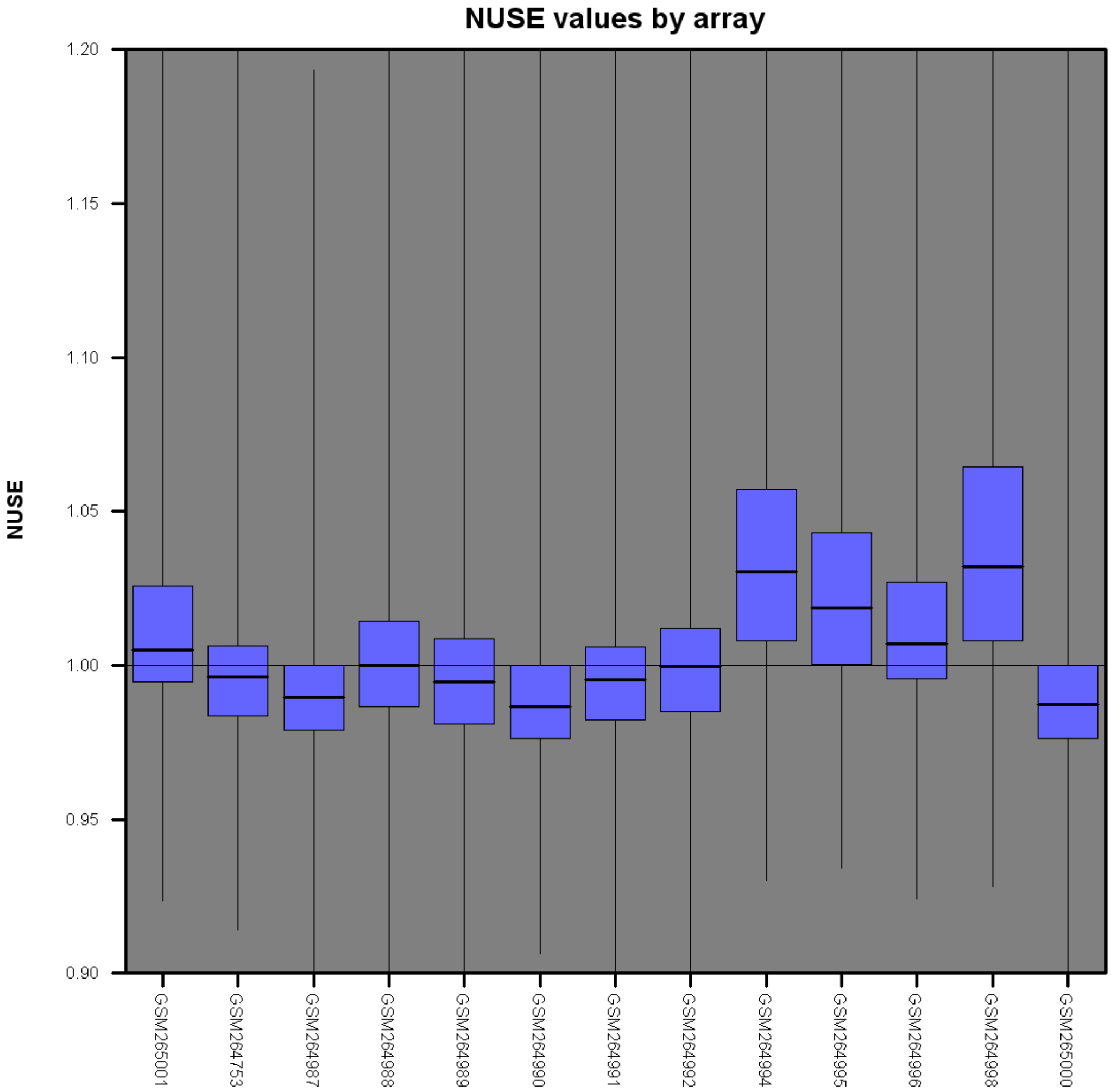

Figure 2-A

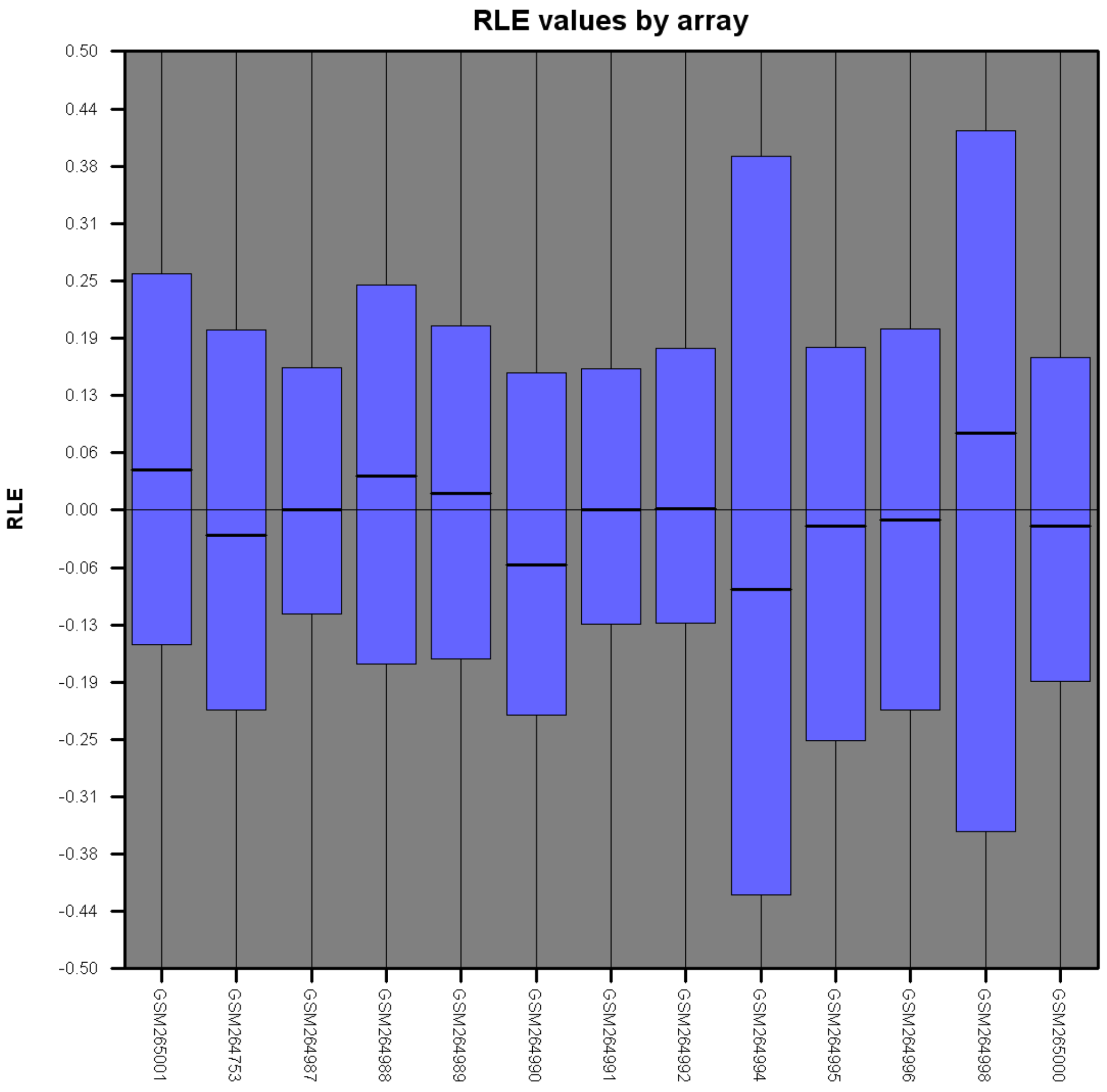

Figure 2-B

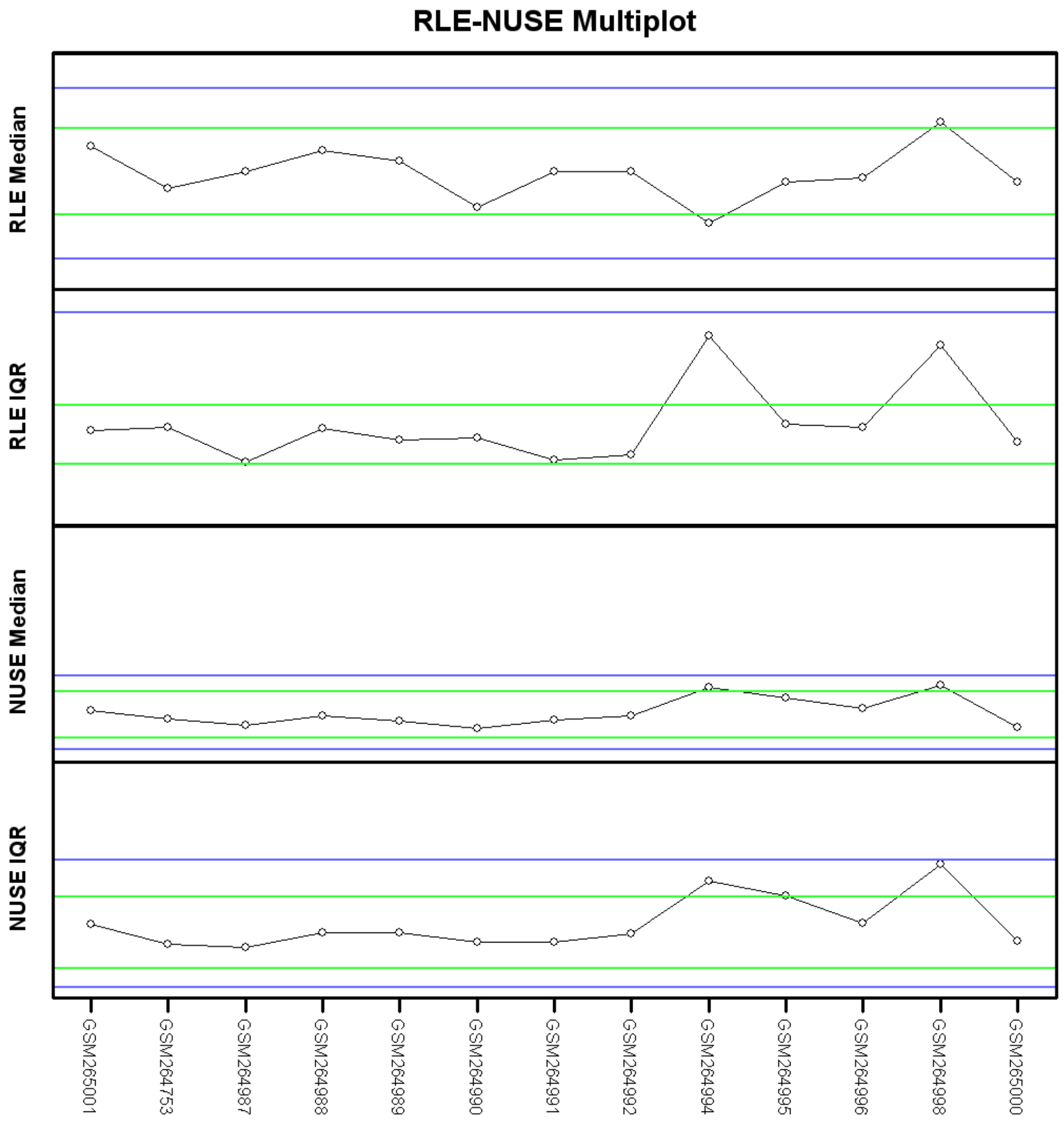

Figure 2-C

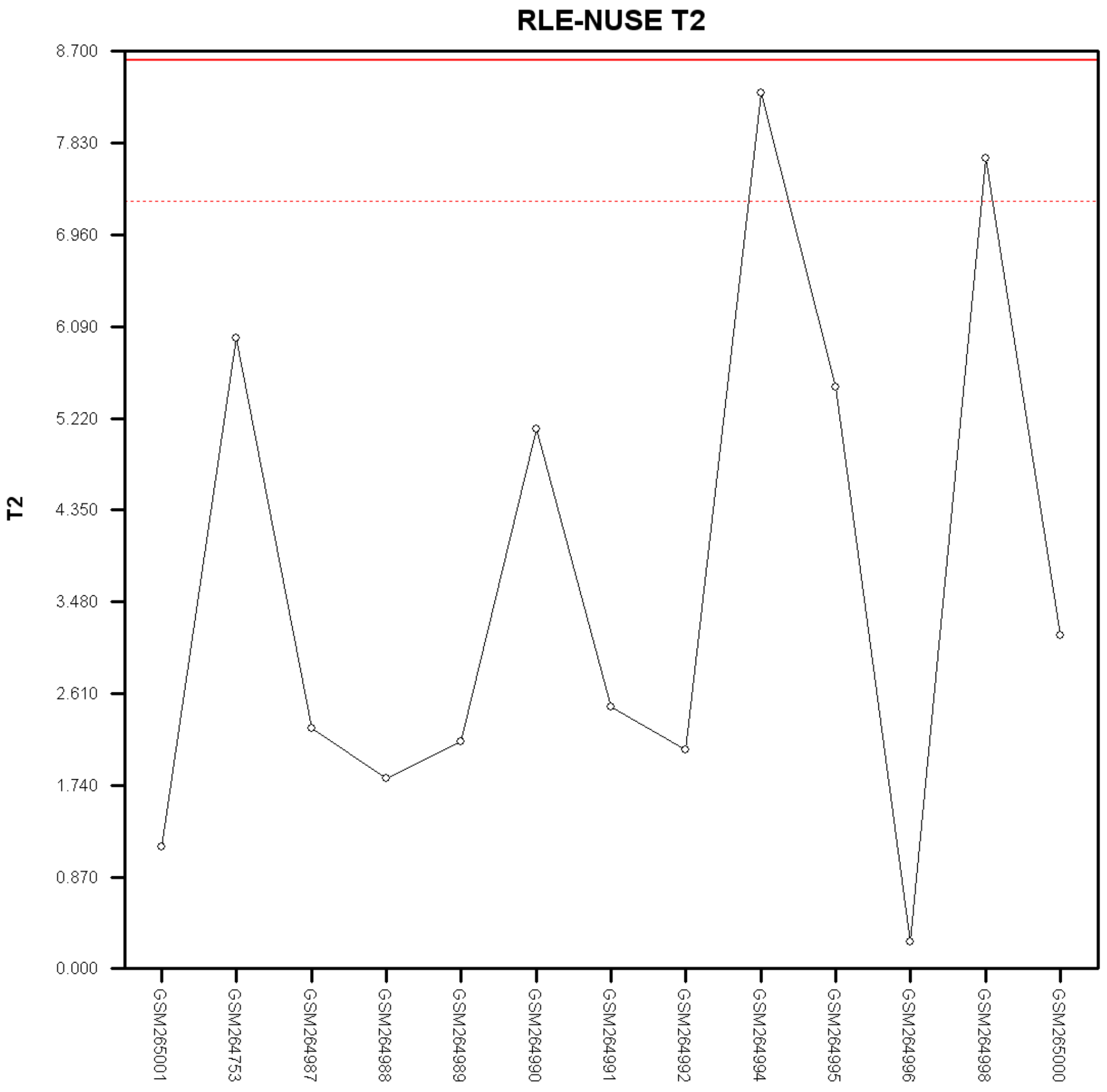

Figure 2-D

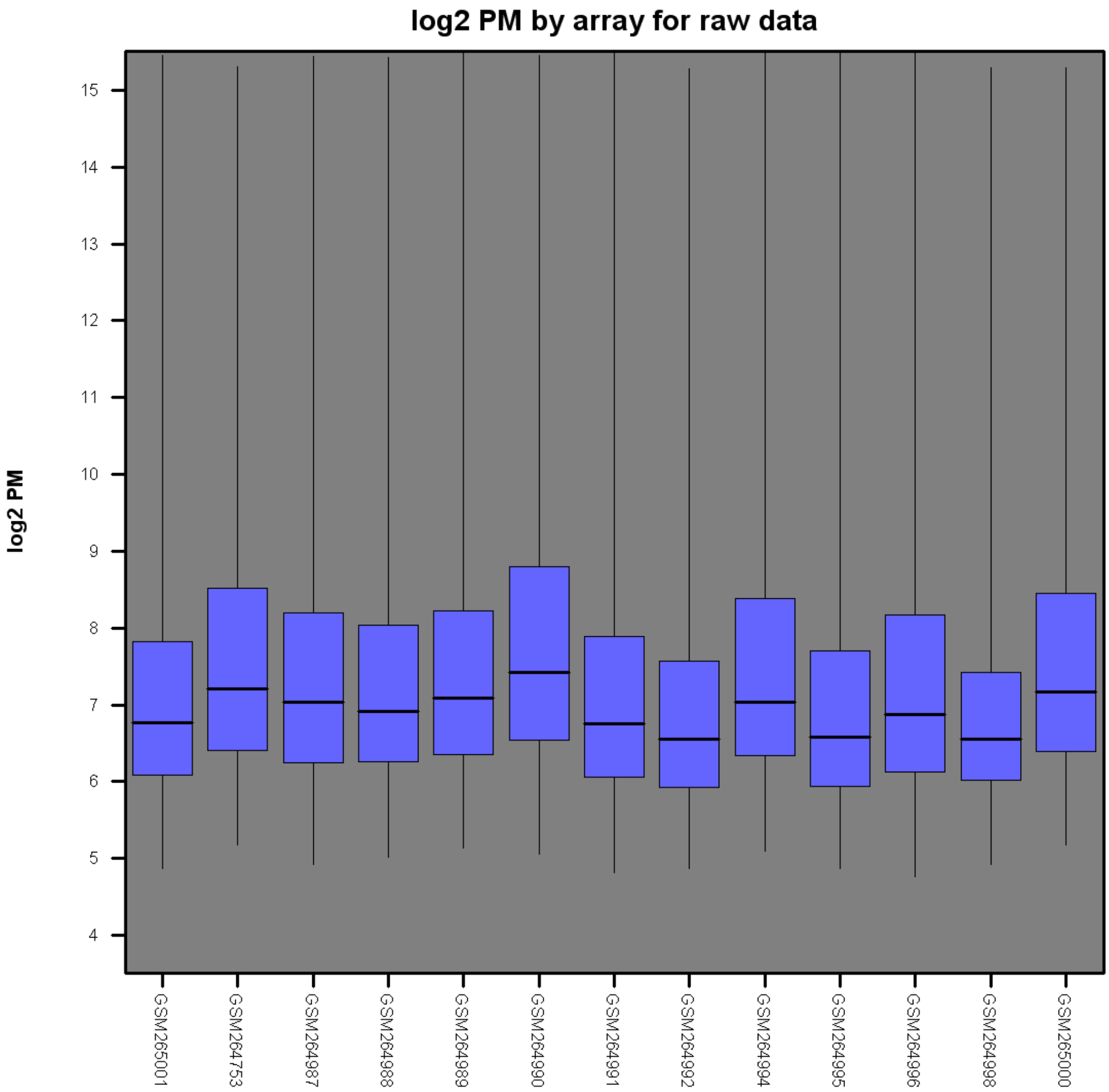

Figure 2-E

Table 1. Toll-like receptor

| Probe Set ID | P value | Arrow | Fold | Gene Symbol |
|---|---|---|---|---|
| 204924_at | 6.89E-09 | up | 3.412607 | TLR2 |
| 210166_at | 1.85E-07 | up | 2.732714 | TLR5 |
| 210176_at | 3.49E-04 | up | 2.256822 | TLR1 |
| 220832_at | 1.23E-07 | up | 5.041227 | TLR8 |
| 221060_s_at | 8.48E-06 | up | 2.713024 | TLR4 |
| 219618_at | 1.79E-09 | up | 3.059633 | IRAK4 |

Table 2. Heat Shock Protein

| Probe Set ID | Pvalue | Arrow | Fold | Gene Symbol |
|---|---|---|---|---|
| 200598_s_at | 5.09E-07 | down | 2.56764 | HSP90B1 |
| 200800_s_at | 8.10E-08 | up | 3.152098 | HSPA1A /// HSPA1B |
| 202557_at | 2.10E-04 | up | 2.438485 | HSPA13 |
| 202581_at | 1.79E-11 | up | 6.14778 | HSPA1A /// HSPA1B |
| 208744_x_at | 1.45E-08 | down | 2.109862 | HSPH1 |
| 208815_x_at | 6.07E-07 | up | 2.042983 | HSPA4 |
| 210338_s_at | 4.69E-08 | down | 2.607109 | HSPA8 |
| 211969_at | 9.22E-14 | down | 15.02711 | HSP90AA1 |
| 219284_at | 4.74E-04 | up | 2.277097 | HSPBAP1 |
| 200941_at | 7.08E-09 | up | 2.341059 | HSBP1 |
| 200942_s_at | 1.40E-05 | up | 2.092604 | HSBP1 |
| 208810_at | 9.82E-04 | up | 2.226605 | DNAJB6 /// TMEM135 |
| 209015_s_at | 2.14E-07 | up | 2.790782 | DNAJB6 |
| 209157_at | 9.23E-10 | up | 3.232302 | DNAJA2 |
| 212467_at | 6.04E-10 | up | 4.209722 | DNAJC13 |
| 212908_at | 1.42E-10 | down | 3.075077 | DNAJC16 |
| 212911_at | 3.41E-09 | up | 3.145019 | DNAJC16 |
| 202842_s_at | 6.90E-04 | up | 2.138169 | DNAJB9 |
| 206782_s_at | 6.78E-09 | up | 2.321206 | DNAJC4 |

Table 3. Chemokine

| Probe Set ID | Pvalue | Arrow | Fold | Gene Symbol |
|---|---|---|---|---|
| 1405_i_at | 5.89E-04 | down | 2.63913 | CCL5 |
| 204103_at | 8.18E-05 | down | 2.515743 | CCL4 |
| 204655_at | 7.65E-04 | down | 2.245439 | CCL5 |
| 205099_s_at | 1.52E-06 | down | 2.461891 | CCR1 |
| 205898_at | 7.09E-06 | down | 4.04859 | CX3CR1 |
| 206337_at | 1.20E-07 | down | 4.466033 | CCR7 |
| 206366_x_at | 5.25E-11 | down | 5.07455 | XCL1 |
| 206991_s_at | 1.51E-04 | down | 2.039673 | CCR5 |
| 208304_at | 4.89E-06 | down | 4.671167 | CCR3 |
| 214567_s_at | 1.62E-08 | down | 3.339736 | XCL1 /// XCL2 |
| 219161_s_at | 4.46E-07 | up | 2.387246 | CKLF |
| 221058_s_at | 9.57E-08 | up | 2.487534 | CKLF |
| 200660_at | 9.31E-10 | up | 2.113736 | S100A11 |
| 200815_s_at | 1.43E-11 | up | 2.871307 | PAFAH1B1 |
| 202917_s_at | 3.16E-09 | up | 2.744783 | S100A8 |
| 203535_at | 2.25E-13 | up | 2.882588 | S100A9 |
| 204351_at | 4.60E-04 | up | 2.44348 | S100P |
| 205863_at | 7.00E-10 | up | 4.3815 | S100A12 |

Table 4. MHC

| Probe Set ID | P value | Arrow | Fold | Gene Symbol |
|---|---|---|---|---|
| 204670_x_at | 9.95E-11 | down | 3.865348 | HLA-DRB1/4 |
| 208306_x_at | 9.05E-09 | down | 2.986262 | HLA-DRB1 |
| 208894_at | 3.15E-10 | down | 4.546559 | HLA-DRA |
| 209312_x_at | 4.77E-09 | down | 3.655453 | HLA-DRB1/4/5 |
| 209823_x_at | 1.81E-04 | down | 2.484667 | HLA-DQB1 |
| 210982_s_at | 8.58E-08 | down | 3.12086 | HLA-DRA |
| 211656_x_at | 4.91E-05 | down | 2.002577 | HLA-DQB1 |
| 211990_at | 1.49E-06 | down | 3.785754 | HLA-DPA1 |
| 211991_s_at | 1.37E-08 | down | 3.178668 | HLA-DPA1 |
| 212671_s_at | 3.46E-04 | down | 2.569759 | HLA-DQA1/2 |
| 212998_x_at | 2.28E-05 | down | 2.456309 | HLA-DQB1 |
| 213537_at | 2.69E-05 | down | 2.602025 | HLA-DPA1 |
| 215193_x_at | 1.62E-08 | down | 3.284869 | HLA-DRB1/3/4 |
| 217478_s_at | 1.25E-07 | down | 2.783175 | HLA-DMA |
| 221491_x_at | 4.58E-06 | down | 2.600531 | HLA-DRB1/3/4/5 |
| 201137_s_at | 1.25E-06 | down | 2.815927 | HLA-DPB1 |
| 203290_at | 4.64E-08 | down | 5.873323 | HLA-DQA1 |
| 203932_at | 7.50E-07 | down | 2.382645 | HLA-DMB |

Table 5. Transcription factor

| Probe Set ID | P value | Arrow | Fold | GeneSymbol |
|---|---|---|---|---|
| 205026_at | 2.14E-10 | up | 2.427908 | STAT5B |
| 208991_at | 5.85E-09 | down | 3.745264 | STAT3 |
| 209969_s_at | 1.14E-05 | down | 3.752185 | STAT1 |
| 212549_at | 4.02E-11 | up | 2.520162 | STAT5B |
| 212550_at | 3.90E-09 | up | 2.643262 | STAT5B |
| 209189_at | 1.67E-04 | up | 2.499937 | FOS |
| 218880_at | 4.95E-08 | up | 3.472536 | FOSL2 |
| 201473_at | 7.18E-08 | up | 2.59816 | JUNB |
| 212501_at | 7.61E-09 | up | 2.240303 | CEBPB |
| 213006_at | 1.41E-09 | up | 3.735119 | CEBPD |
| 214523_at | 2.02E-06 | up | 2.091157 | CEBPE |
| 204039_at | 1.40E-08 | up | 2.398913 | CEBPA |
| 204203_at | 1.31E-08 | up | 2.358698 | CEBPG |
| 203574_at | 1.13E-08 | up | 4.640286 | NFIL3 |
| 201502_s_at | 5.83E-06 | down | 2.115988 | NFKBIA |
| 205841_at | 2.40E-13 | up | 5.992293 | JAK2 |
| 205842_s_at | 2.73E-06 | up | 3.288805 | JAK2 |
| 209604_s_at | 4.81E-16 | down | 6.909352 | GATA3 |
| 210555_s_at | 1.03E-05 | down | 2.547481 | NFATC3 |
| 210556_at | 5.20E-05 | down | 2.429433 | NFATC3 |
| 215092_s_at | 1.12E-05 | down | 2.197351 | NFAT5 |
| 217526_at | 3.36E-08 | down | 3.459827 | NFATC2IP |
| 217527_s_at | 1.86E-10 | down | 4.586211 | NFATC2IP |
| 10426_x_at | 8.86E-10 | down | 5.364871 | RORA |
| 210479_s_at | 9.23E-10 | down | 6.440884 | RORA |

Table 6. Leukotriene & prostaglandin

| Probe Set ID | P value | Arrow | Fold | GeneSymbol |
|---|---|---|---|---|
| 208771_s_at | 1.53E-08 | up | 2.726662 | LTA4H |
| 210128_s_at | 8.83E-10 | up | 3.067644 | LTB4R |
| 216388_s_at | 4.30E-09 | up | 2.518544 | LTB4R |
| 215894_at | 1.13E-11 | down | 9.422997 | PTGDR |
| 203913_s_at | 3.16E-10 | up | 27.23115 | HPGD |
| 203914_x_at | 3.10E-09 | up | 19.87965 | HPGD |
| 204445_s_at | 1.10E-06 | up | 2.227303 | ALOX5 |
| 204446_s_at | 3.10E-08 | up | 2.322821 | ALOX5 |
| 204614_at | 3.94E-06 | up | 2.905599 | SERPINB2 |
| 209533_s_at | 1.09E-08 | up | 2.612623 | PLAA |
| 210145_at | 1.31E-09 | up | 3.562516 | PLA2G4A |
| 210772_at | 3.05E-08 | up | 4.401296 | FPR2 |
| 210773_s_at | 3.42E-06 | up | 3.867929 | FPR2 |
| 213572_s_at | 2.12E-12 | up | 6.133628 | SERPINB1 |
| 214366_s_at | 4.55E-09 | up | 3.975351 | ALOX5 |

Table7. MMP and FGF

| Probe Set ID | Pvalue | Arrow | Fold | Gene Symbol |
|---|---|---|---|---|
| 207329_at | 1.17E-08 | up | 28.02386 | MMP8 |
| 207890_s_at | 1.46E-09 | up | 3.310085 | MMP25 |
| 203936_s_at | 2.84E-12 | up | 11.50853 | MMP9 |
| 205110_s_at | 2.26E-06 | up | 5.236618 | FGF13 |
| 201666_at | 1.46E-08 | up | 2.519345 | TIMP1 |
| 203167_at | 2.47E-10 | up | 3.173297 | TIMP2 |
| 219295_s_at | 7.61E-07 | up | 6.184153 | PCOLCE2 |
| 200827_at | 2.27E-07 | up | 2.158958 | PLOD1 |
| 200654_at | 1.47E-09 | up | 2.244832 | P4HB |
| 201940_at | 2.24E-08 | up | 5.123088 | CPD |
| 201941_at | 8.36E-08 | up | 4.763787 | CPD |
| 201942_s_at | 2.76E-06 | up | 3.149759 | CPD |
| 201943_s_at | 6.79E-10 | up | 6.419939 | CPD |
| 202304_at | 1.23E-09 | up | 3.617562 | FNDC3A |
| 203044_at | 1.34E-05 | up | 3.342165 | CHSY1 |
| 203284_s_at | 2.64E-08 | up | 3.515453 | HS2ST1 |
| 203285_s_at | 9.15E-10 | up | 2.626396 | HS2ST1 |
| 207165_at | 7.94E-05 | up | 2.063695 | HMMR |
| 207543_s_at | 9.43E-07 | up | 2.718742 | P4HA1 |
| 211945_s_at | 0.001738 | up | 2.065411 | ITGB1 |
| 218718_at | 2.24E-10 | up | 12.04417 | PDGFC |
| 219049_at | 1.52E-08 | up | 7.475137 | CSGALNACT1 |
| 219403_s_at | 1.33E-07 | up | 5.223431 | HPSE |
| 222235_s_at | 3.83E-10 | up | 13.31196 | CSGALNACT2 |

Table 8. Complement

| Probe Set ID | Pvalue | Arrow | Fold | Gene Symbol |
|---|---|---|---|---|
| 200983_x_at | 1.18E-09 | up | 4.196889 | CD59 |
| 200984_s_at | 1.67E-10 | up | 4.910066 | CD59 |
| 200985_s_at | 3.02E-11 | up | 8.311746 | CD59 |
| 201925_s_at | 3.14E-07 | up | 6.090309 | CD55 |
| 201926_s_at | 4.95E-09 | up | 4.097339 | CD55 |
| 202953_at | 5.10E-04 | up | 2.016919 | C1QB |
| 205786_s_at | 9.34E-11 | up | 3.896006 | ITGAM |
| 206244_at | 5.96E-11 | up | 7.560091 | CR1 |
| 209906_at | 5.21E-09 | up | 5.687038 | C3AR1 |
| 210184_at | 2.03E-05 | up | 2.185833 | ITGAX |
| 212463_at | 3.67E-08 | up | 3.248518 | CD59 |
| 217552_x_at | 1.83E-09 | up | 3.938783 | CR1 |
| 218232_at | 2.16E-05 | up | 3.030927 | C1QA |
| 218983_at | 2.04E-07 | up | 3.029844 | C1RL |
| 220088_at | 2.45E-06 | up | 2.500003 | C5AR1 |
| 205033_s_at | 7.03E-06 | up | 6.365769 | DEFA1/1B/3 |
| 207269_at | 1.49E-04 | up | 6.195451 | DEFA4 |

Table 9. Cathepsin

| Probe Set ID | Pvalue | Arrow | Fold | Gene Symbol |
|---|---|---|---|---|
| 202450_s_at | 4.73E-06 | up | 2.020977 | CTSK |
| 203758_at | 3.84E-07 | down | 2.476479 | CTSO |
| 205653_at | 4.28E-04 | up | 3.634934 | CTSG |
| 210042_s_at | 5.41E-05 | up | 2.824491 | CTSZ |
| 214450_at | 9.49E-04 | down | 2.150796 | CTSW |
| 200661_at | 1.10E-07 | up | 2.603046 | CTSA |
| 200766_at | 3.02E-11 | up | 3.793397 | CTSD |
| 201487_at | 2.22E-07 | up | 3.024736 | CTSC |
| 203948_s_at | 6.35E-04 | up | 2.112642 | MPO |
| 203949_at | 5.14E-06 | up | 4.61238 | MPO |
| 204961_s_at | 5.23E-06 | up | 2.092114 | NCF1B1C |
| 207677_s_at | 5.21E-09 | up | 3.07743 | NCF4 |

Table10. CSF

| Probe Set ID | Pvalue | Arrow | Fold | Gene Symbol |
|---|---|---|---|---|
| 205159_at | 7.47E-05 | up | 2.272558 | CSF2RB |
| 210340_s_at | 6.06E-10 | up | 2.727757 | CSF2RA |
| 203591_s_at | 4.27E-06 | up | 2.365631 | CSF3R |

Table 11. Fc receptor

| Probe Set ID | Pvalue | Arrow | Fold | Gene Symbol |
|---|---|---|---|---|
| 203561_at | 5.57E-08 | up | 2.06512 | FCGR2A |
| 204232_at | 6.41E-10 | up | 2.89912 | FCER1G |
| 207674_at | 4.21E-07 | up | 6.511793 | FCAR |
| 210992_x_at | 2.25E-05 | up | 2.003132 | FCGR2C |
| 211307_s_at | 3.73E-07 | up | 4.462972 | FCAR |
| 211734_s_at | 7.11E-05 | down | 4.276542 | FCER1A |
| 211816_x_at | 2.46E-05 | up | 2.47651 | FCAR |
| 214511_x_at | 1.06E-05 | up | 3.171251 | FCGR1B |
| 216950_s_at | 4.67E-08 | up | 5.148257 | FCGR1A/1C |

Table 12. Cytokine & receptor

| Probe Set ID | Pvalue | Arrow | Fold | Gene Symbol |
|---|---|---|---|---|
| 203828_s_at | 6.11E-04 | down | 2.216758 | IL32 |
| 205227_at | 4.73E-04 | up | 2.365252 | IL1RAP |
| 205291_at | 2.41E-06 | down | 3.160494 | IL2RB |
| 205403_at | 5.96E-11 | up | 9.990063 | IL1R2 |
| 205707_at | 5.70E-06 | down | 2.016105 | IL17RA |
| 205798_at | 2.19E-21 | down | 28.62358 | IL7R |
| 205926_at | 2.78E-10 | down | 2.211585 | IL27RA |
| 205945_at | 9.53E-14 | down | 13.61186 | IL6R |
| 205992_s_at | 3.86E-06 | up | 3.345045 | IL15 |
| 206618_at | 1.05E-11 | up | 17.52686 | IL18R1 |
| 207072_at | 3.95E-11 | up | 6.322352 | IL18RAP |
| 208200_at | 3.63E-09 | down | 4.584162 | IL1A |
| 208930_s_at | 7.16E-10 | down | 4.59278 | ILF3 |
| 211372_s_at | 2.35E-10 | up | 17.05508 | IL1R2 |
| 212195_at | 9.18E-05 | up | 2.753398 | IL6ST |
| 212196_at | 1.44E-05 | up | 2.060642 | IL6ST |
| 212657_s_at | 3.37E-06 | up | 2.343125 | IL1RN |
| 217489_s_at | 9.59E-13 | down | 3.415206 | IL6R |
| 202948_at | 1.06E-10 | up | 9.925212 | IL1R1 |
| 203233_at | 1.18E-09 | up | 3.541053 | IL4R |
| 205016_at | 6.86E-09 | up | 4.867131 | TGFA |
| 201506_at | 1.41E-04 | down | 2.289291 | TGFBI |
| 203085_s_at | 1.35E-05 | up | 2.13325 | TGFB1 |
| 204731_at | 3.89E-15 | down | 10.80744 | TGFBR3 |
| 206026_s_at | 7.23E-06 | up | 4.685213 | TNFAIP6 |
| 206222_at | 3.55E-07 | down | 2.189329 | TNFRSF10C |
| 207536_s_at | 4.78E-06 | down | 2.923358 | TNFRSF9 |
| 207643_s_at | 1.78E-08 | up | 2.618468 | TNFRSF1A |
| 207907_at | 3.57E-08 | down | 3.31319 | TNFSF14 |
| 208296_x_at | 8.56E-05 | up | 2.646926 | TNFAIP8 |
| 210260_s_at | 5.80E-05 | up | 2.88896 | TNFAIP8 |
| 214329_x_at | 2.56E-04 | up | 2.679365 | TNFSF10 |
| 202509_s_at | 8.86E-10 | down | 2.258532 | TNFAIP2 |
| 208114_s_at | 6.33E-16 | down | 6.286403 | ISG20L2 |
| 208965_s_at | 1.52E-08 | down | 6.302218 | IFI16 |
| 211676_s_at | 1.42E-07 | up | 4.556334 | IFNGR1 |
| 220577_at | 8.19E-07 | down | 2.331832 | GVINP1 |
| 201642_at | 8.60E-08 | up | 2.225393 | IFNGR2 |
| 202269_x_at | 1.73E-05 | down | 4.987527 | GBP1 |
| 202727_s_at | 9.68E-08 | up | 3.452631 | IFNGR1 |
| 204191_at | 2.33E-07 | up | 2.031339 | IFNAR1 |
| 204415_at | 0.004538 | up | 2.774495 | IFI6 |
| 204439_at | 0.004491 | down | 3.991523 | IFI44L |
| 204747_at | 5.65E-04 | down | 3.737294 | IFIT3 |
| 204786_s_at | 5.62E-17 | down | 6.491528 | IFNAR2 |
| 200704_at | 1.66E-07 | up | 2.056096 | LITAF |
| 201108_s_at | 3.76E-05 | up | 2.314963 | THBS1 |
| 201109_s_at | 2.60E-05 | up | 3.128824 | THBS1 |
| 201110_s_at | 1.47E-08 | up | 7.142229 | THBS1 |
| 204780_s_at | 3.86E-04 | up | 2.720255 | FAS |
| 204781_s_at | 9.11E-05 | up | 2.032212 | FAS |
| 221601_s_at | 1.46E-09 | down | 4.684259 | FAIM3 |
| 221602_s_at | 2.90E-10 | down | 3.944784 | FAIM3 |

Table 13. CD molecules

| Probe Set ID | Pvalue | Arrow | Fold | Gene Symbol |
|---|---|---|---|---|
| 200663_at | 1.23E-10 | up | 2.64625 | CD63 |
| 201005_at | 1.19E-06 | up | 4.053223 | CD9 |
| 202878_s_at | 1.20E-04 | up | 2.316249 | CD93 |
| 202910_s_at | 1.39E-04 | up | 2.078598 | CD97 |
| 203645_s_at | 1.60E-06 | up | 7.818829 | CD163 |
| 203799_at | 0.002884 | up | 2.038077 | CD302 |
| 204489_s_at | 3.25E-10 | up | 2.475646 | CD44 |
| 204490_s_at | 1.31E-08 | up | 2.582747 | CD44 |
| 204627_s_at | 3.45E-05 | up | 3.667203 | ITGB3 |
| 204661_at | 7.10E-06 | down | 2.483425 | CD52 |
| 205173_x_at | 7.28E-08 | up | 4.151193 | CD58 |
| 205758_at | 1.12E-04 | down | 3.211919 | CD8A |
| 205789_at | 4.22E-05 | up | 2.844588 | CD1D |
| 205831_at | 2.21E-07 | down | 4.421892 | CD2 |
| 205987_at | 1.13E-07 | down | 2.078875 | CD1C |
| 205988_at | 1.13E-15 | down | 5.71907 | CD84 |
| 206150_at | 2.30E-08 | down | 2.506513 | CD27 |
| 206488_s_at | 6.10E-04 | up | 2.481129 | CD36 |
| 206493_at | 5.05E-06 | up | 3.000028 | ITGA2B |
| 206494_s_at | 7.02E-04 | up | 3.137038 | ITGA2B |
| 206761_at | 3.30E-04 | down | 2.113857 | CD96 |
| 206804_at | 6.16E-10 | down | 4.490583 | CD3G |
| 208405_s_at | 8.52E-05 | up | 2.330166 | CD164 |
| 208650_s_at | 7.09E-07 | up | 6.548884 | CD24 |
| 208651_x_at | 6.39E-08 | up | 4.732706 | CD24 |
| 208652_at | 1.21E-07 | up | 2.66052 | PPP2CA |
| 208653_s_at | 6.26E-10 | up | 5.120652 | CD164 |
| 208654_s_at | 3.97E-06 | up | 5.608513 | CD164 |
| 209555_s_at | 2.95E-04 | up | 2.950439 | CD36 |
| 209771_x_at | 1.32E-07 | up | 6.559978 | CD24 |
| 209835_x_at | 1.06E-06 | up | 2.202673 | CD44 |
| 210031_at | 1.34E-06 | down | 3.413696 | CD247 |
| 210184_at | 2.03E-05 | up | 2.185833 | ITGAX |
| 210895_s_at | 2.39E-04 | down | 2.251445 | CD86 |
| 211744_s_at | 5.92E-08 | up | 4.172478 | CD58 |
| 211893_x_at | 8.92E-10 | down | 2.082775 | CD6 |
| 211900_x_at | 5.14E-12 | down | 2.56049 | CD6 |
| 212014_x_at | 7.54E-07 | up | 2.308336 | CD44 |
| 212063_at | 4.06E-06 | up | 2.119029 | CD44 |
| 213958_at | 1.17E-06 | down | 2.205519 | CD6 |
| 215049_x_at | 2.45E-06 | up | 7.967604 | CD163 |
| 215240_at | 1.19E-08 | up | 2.467146 | ITGB3 |
| 216233_at | 4.99E-06 | up | 6.556412 | CD163 |
| 216331_at | 0.001993 | up | 2.434621 | ITGA7 |
| 216379_x_at | 5.86E-08 | up | 7.598393 | CD24 |
| 216942_s_at | 1.49E-05 | up | 3.193467 | CD58 |
| 216956_s_at | 1.65E-04 | up | 2.29709 | ITGA2B |
| 217523_at | 9.19E-10 | down | 6.259623 | CD44 |
| 219669_at | 4.80E-13 | up | 52.83338 | CD177 |
| 222061_at | 2.36E-09 | up | 3.870316 | CD58 |
| 266_s_at | 6.15E-09 | up | 9.687818 | CD24 |
| 213539_at | 1.57E-06 | down | 3.240318 | CD3D |

Table 14. NK/CTL molecules

| Probe Set ID | Pvalue | Arrow | Fold | Gene Symbol |
|---|---|---|---|---|
| 205821_at | 8.68E-07 | down | 3.717665 | KLRK1 |
| 206666_at | 6.47E-06 | down | 3.898898 | GZMK |
| 207460_at | 5.21E-06 | down | 2.196101 | GZMM |
| 207795_s_at | 8.11E-05 | down | 2.646551 | KLRD1 |
| 210164_at | 3.84E-06 | down | 4.671744 | GZMB |
| 210288_at | 2.27E-09 | down | 5.673069 | KLRG1 |
| 210321_at | 2.14E-05 | down | 6.098404 | GZMH |
| 210606_x_at | 2.18E-05 | down | 2.995347 | KLRD1 |
| 210915_x_at | 2.72E-06 | down | 3.166231 | TRBC1 |
| 210972_x_at | 3.84E-07 | down | 3.223779 | TRAC/J17/V20 |
| 211796_s_at | 2.86E-06 | down | 3.233982 | TRBC1/C2 |
| 211902_x_at | 3.75E-06 | down | 2.651879 | TRD@ |
| 213193_x_at | 8.35E-07 | down | 3.394675 | TRBC1 |
| 213830_at | 1.16E-08 | down | 3.694018 | TRD@ |
| 214470_at | 4.30E-04 | down | 2.832583 | KLRB1 |
| 214617_at | 2.56E-04 | down | 3.489038 | PRF1 |
| 215338_s_at | 2.44E-19 | down | 13.36511 | NKTR |
| 215806_x_at | 2.27E-05 | down | 3.907998 | TARP/TRGC2 |
| 216191_s_at | 3.85E-06 | down | 5.215065 | TRDV3 |
| 216920_s_at | 1.31E-06 | down | 5.017736 | TARP/TRGC2 |
| 217143_s_at | 6.84E-08 | down | 6.090665 | TRD@ |
| 220646_s_at | 0.004097 | down | 2.571194 | KLRF1 |
| 37145_at | 4.61E-06 | down | 5.027259 | GNLY |

## References

1. Kong SL, Chui P, Lim B, Salto-Tellez M. Elucidating the molecular physiopathology of acute respiratory distress syndrome in severe acute respiratory syndrome patients. *Virus Res* 2009;145:260-269.
2. Lee EJ, Lim JY, Lee SY, Lee SH, In KH, Yoo SH, Sul D, Park S. The expression of hsps, anti-oxidants, and cytokines in plasma and bronchoalveolar lavage fluid of patients with acute respiratory distress syndrome. *Clin Biochem* 2012;45:493-498.
3. Papadakos PJ. Cytokines, genes, and ards. *Chest* 2002;121:1391-1392.
4. Downey GP, Dong Q, Kruger J, Dedhar S, Cherapanov V. Regulation of neutrophil activation in acute lung injury. *Chest* 1999;116:46S-54S.
5. Lee WL, Downey GP. Neutrophil activation and acute lung injury. *Curr Opin Crit Care* 2001;7:1-7.
6. Howrylak JA, Dolinay T, Lucht L, Wang Z, Christiani DC, Sethi JM, Xing EP, Donahoe MP, Choi AM. Discovery of the gene signature for acute lung injury in patients with sepsis. *Physiol Genomics* 2009;37:133-139.
7. Rotunno M, Hu N, Su H, Wang C, Goldstein AM, Bergen AW, Consonni D, Pesatori AC, Bertazzi PA, Wacholder S, Shih J, Caporaso NE, Taylor PR, Landi MT. A gene expression signature from peripheral whole blood for stage i lung adenocarcinoma. *Cancer Prev Res (Phila)* 2011;4:1599-1608.
8. Cohen AC, Nadeau KC, Tu W, Hwa V, Dionis K, Bezrodnik L, Teper A, Gaillard M, Heinrich J, Krensky AM, Rosenfeld RG, Lewis DB. Cutting edge: Decreased accumulation and regulatory function of cd4+ cd25(high) t cells in human stat5b deficiency. *J Immunol* 2006;177:2770-2774.
9. Yang XO, Pappu BP, Nurieva R, Akimzhanov A, Kang HS, Chung Y, Ma L, Shah B, Panopoulos AD, Schluns KS, Watowich SS, Tian Q, Jetten AM, Dong C. T helper 17 lineage differentiation is programmed by orphan nuclear receptors ror alpha and ror gamma. *Immunity* 2008;28:29-39.
10. Kong MY, Li Y, Oster R, Gaggar A, Clancy JP. Early elevation of matrix metalloproteinase-8 and -9 in pediatric ards is associated with an increased risk of prolonged mechanical ventilation. *PLoS One* 2011;6:e22596.
11. Venkatesan N, Ouzzine M, Kolb M, Netter P, Ludwig MS. Increased deposition of chondroitin/dermatan sulfate glycosaminoglycan and upregulation of beta1,3-glucuronosyltransferase i in pulmonary fibrosis. *Am J Physiol Lung Cell Mol Physiol* 2011;300:L191-203.
12. Hadkar V, Sangsree S, Vogel SM, Brovkovych V, Skidgel RA. Carboxypeptidase-mediated enhancement of nitric oxide production in rat lungs and microvascular endothelial cells. *Am J Physiol Lung Cell Mol Physiol* 2004;287:L35-45.
13. Kobayashi A, Hashimoto S, Kooguchi K, Kitamura Y, Onodera H, Urata Y, Ashihara

T. Expression of inducible nitric oxide synthase and inflammatory cytokines in alveolar macrophages of ards following sepsis. *Chest* 1998;113:1632-1639.
14. Escobar Galvis ML, Jia J, Zhang X, Jastrebova N, Spillmann D, Gottfridsson E, van Kuppevelt TH, Zcharia E, Vlodavsky I, Lindahl U, Li JP. Transgenic or tumor-induced expression of heparanase upregulates sulfation of heparan sulfate. *Nat Chem Biol* 2007;3:773-778.
15. Li J, Shworak NW, Simons M. Increased responsiveness of hypoxic endothelial cells to fgf2 is mediated by hif-1alpha-dependent regulation of enzymes involved in synthesis of heparan sulfate fgf2-binding sites. *J Cell Sci* 2002;115:1951-1959.
16. Myllyharju J. Prolyl 4-hydroxylases, key enzymes in the synthesis of collagens and regulation of the response to hypoxia, and their roles as treatment targets. *Ann Med* 2008;40:402-417.
17. Turpeenniemi-Hujanen TM. Immunological characterization of lysyl hydroxylase, an enzyme of collagen synthesis. *Biochem J* 1981;195:669-676.
18. Charles PE, Tissieres P, Barbar SD, Croisier D, Dufour J, Dunn-Siegrist I, Chavanet P, Pugin J. Mild-stretch mechanical ventilation upregulates toll-like receptor 2 and sensitizes the lung to bacterial lipopeptide. *Crit Care* 2011;15:R181.
19. Fan J, Li Y, Vodovotz Y, Billiar TR, Wilson MA. Hemorrhagic shock-activated neutrophils augment tlr4 signaling-induced tlr2 upregulation in alveolar macrophages: Role in hemorrhage-primed lung inflammation. *Am J Physiol Lung Cell Mol Physiol* 2006;290:L738-L746.
20. Imai Y, Kuba K, Neely GG, Yaghubian-Malhami R, Perkmann T, van Loo G, Ermolaeva M, Veldhuizen R, Leung YH, Wang H, Liu H, Sun Y, Pasparakis M, Kopf M, Mech C, Bavari S, Peiris JS, Slutsky AS, Akira S, Hultqvist M, Holmdahl R, Nicholls J, Jiang C, Binder CJ, Penninger JM. Identification of oxidative stress and toll-like receptor 4 signaling as a key pathway of acute lung injury. *Cell* 2008;133:235-249.
21. Jiang D, Liang J, Fan J, Yu S, Chen S, Luo Y, Prestwich GD, Mascarenhas MM, Garg HG, Quinn DA, Homer RJ, Goldstein DR, Bucala R, Lee PJ, Medzhitov R, Noble PW. Regulation of lung injury and repair by toll-like receptors and hyaluronan. *Nat Med* 2005;11:1173-1179.
22. Lv T, Shen X, Shi Y, Song Y. Tlr4 is essential in acute lung injury induced by unresuscitated hemorrhagic shock. *J Trauma* 2009;66:124-131.
23. Sharif R, Dawra R, Wasiluk K, Phillips P, Dudeja V, Kurt-Jones E, Finberg R, Saluja A. Impact of toll-like receptor 4 on the severity of acute pancreatitis and pancreatitis-associated lung injury in mice. *Gut* 2009;58:813-819.
24. Togbe D, Schnyder-Candrian S, Schnyder B, Couillin I, Maillet I, Bihl F, Malo D, Ryffel B, Quesniaux VF. Tlr4 gene dosage contributes to endotoxin-induced acute respiratory inflammation. *J Leukoc Biol* 2006;80:451-457.

25. Wu TT, Chen TL, Loon WS, Tai YT, Cherng YG, Chen RM. Lipopolysaccharide stimulates syntheses of toll-like receptor 2 and surfactant protein-a in human alveolar epithelial a549 cells through upregulating phosphorylation of mek1 and erk1/2 and sequential activation of nf-kappab. *Cytokine* 2011;55:40-47.
26. Reino DC, Pisarenko V, Palange D, Doucet D, Bonitz RP, Lu Q, Colorado I, Sheth SU, Chandler B, Kannan KB, Ramanathan M, Xu da Z, Deitch EA, Feinman R. Trauma hemorrhagic shock-induced lung injury involves a gut-lymph-induced tlr4 pathway in mice. *PLoS One* 2011;6:e14829.
27. Chase MA, Wheeler DS, Lierl KM, Hughes VS, Wong HR, Page K. Hsp72 induces inflammation and regulates cytokine production in airway epithelium through a tlr4- and nf-kappab-dependent mechanism. *J Immunol* 2007;179:6318-6324.
28. Ganter MT, Ware LB, Howard M, Roux J, Gartland B, Matthay MA, Fleshner M, Pittet JF. Extracellular heat shock protein 72 is a marker of the stress protein response in acute lung injury. *Am J Physiol Lung Cell Mol Physiol* 2006;291:L354-361.
29. Krzyzaniak M, Cheadle G, Peterson C, Loomis W, Putnam J, Wolf P, Baird A, Eliceiri B, Bansal V, Coimbra R. Burn-induced acute lung injury requires a functional toll-like receptor 4. *Shock* 2011;36:24-29.
30. Khair OA, Davies RJ, Devalia JL. Bacterial-induced release of inflammatory mediators by bronchial epithelial cells. *Eur Respir J* 1996;9:1913-1922.
31. Smart SJ, Casale TB. Pulmonary epithelial cells facilitate tnf-alpha-induced neutrophil chemotaxis. A role for cytokine networking. *J Immunol* 1994;152:4087-4094.
32. Thorley AJ, Ford PA, Giembycz MA, Goldstraw P, Young A, Tetley TD. Differential regulation of cytokine release and leukocyte migration by lipopolysaccharide-stimulated primary human lung alveolar type ii epithelial cells and macrophages. *J Immunol* 2007;178:463-473.
33. Kurdowska A, Miller EJ, Noble JM, Baughman RP, Matthay MA, Brelsford WG, Cohen AB. Anti-il-8 autoantibodies in alveolar fluid from patients with the adult respiratory distress syndrome. *J Immunol* 1996;157:2699-2706.
34. Kunkel SL, Standiford T, Kasahara K, Strieter RM. Interleukin-8 (il-8): The major neutrophil chemotactic factor in the lung. *Exp Lung Res* 1991;17:17-23.
35. Lin G, Pearson AE, Scamurra RW, Zhou Y, Baarsch MJ, Weiss DJ, Murtaugh MP. Regulation of interleukin-8 expression in porcine alveolar macrophages by bacterial lipopolysaccharide. *J Biol Chem* 1994;269:77-85.
36. Nakamura H, Yoshimura K, Jaffe HA, Crystal RG. Interleukin-8 gene expression in human bronchial epithelial cells. *J Biol Chem* 1991;266:19611-19617.
37. Standiford TJ, Kunkel SL, Basha MA, Chensue SW, Lynch JP, 3rd, Toews GB, Westwick J, Strieter RM. Interleukin-8 gene expression by a pulmonary epithelial cell line. A model for cytokine networks in the lung. *J Clin Invest* 1990;86:1945-1953.
38. Strieter RM, Chensue SW, Basha MA, Standiford TJ, Lynch JP, Baggiolini M, Kunkel SL. Human alveolar macrophage gene expression of interleukin-8 by tumor necrosis factor-alpha, lipopolysaccharide, and interleukin-1 beta. *Am J Respir Cell Mol Biol* 1990;2:321-326.
39. De Luca D, Minucci A, Cogo P, Capoluongo ED, Conti G, Pietrini D, Carnielli VP, Piastra M. Secretory phospholipase a pathway during pediatric acute respiratory distress syndrome: A preliminary study. *Pediatr Crit Care Med* 2011;12:e20-24.
40. Donnelly SC, Strieter RM, Kunkel SL, Walz A, Robertson CR, Carter DC, Grant IS, Pollok AJ, Haslett C. Interleukin-8 and development of adult respiratory distress syndrome in at-risk patient groups. *Lancet* 1993;341:643-647.
41. Suzuki K, Miyasaka H, Ota H, Yamakawa Y, Tagawa M, Kuramoto A, Mizuno S. Purification and partial primary sequence of a chemotactic protein for polymorphonuclear leukocytes derived from human lung giant cell carcinoma lu65c cells. *J Exp Med* 1989;169:1895-1901.
42. Frevert CW, Goodman RB, Kinsella MG, Kajikawa O, Ballman K, Clark-Lewis I, Proudfoot AE, Wells TN, Martin TR. Tissue-specific mechanisms control the retention of il-8 in lungs and skin. *J Immunol* 2002;168:3550-3556.
43. Frevert CW, Kinsella MG, Vathanaprida C, Goodman RB, Baskin DG, Proudfoot A, Wells TN, Wight TN, Martin TR. Binding of interleukin-8 to heparan sulfate and chondroitin sulfate in lung tissue. *Am J Respir Cell Mol Biol* 2003;28:464-472.
44. Takatsuka H, Takemoto Y, Mori A, Okamoto T, Kanamaru A, Kakishita E. Common features in the onset of ards after administration of granulocyte colony-stimulating factor. *Chest* 2002;121:1716-1720.
45. Tanaka S, Nishiumi S, Nishida M, Mizushina Y, Kobayashi K, Masuda A, Fujita T, Morita Y, Mizuno S, Kutsumi H, Azuma T, Yoshida M. Vitamin k3 attenuates lipopolysaccharide-induced acute lung injury through inhibition of nuclear factor-kappab activation. *Clin Exp Immunol* 2010;160:283-292.
46. von Bismarck P, Klemm K, Garcia Wistadt CF, Winoto-Morbach S, Schutze S, Krause MF. Selective nf-kappab inhibition, but not dexamethasone, decreases acute lung injury in a newborn piglet airway inflammation model. *Pulm Pharmacol Ther* 2009;22:297-304.
47. Wu CL, Lin LY, Yang JS, Chan MC, Hsueh CM. Attenuation of lipopolysaccharide-induced acute lung injury by treatment with il-10. *Respirology* 2009;14:511-521.
48. Van den Steen PE, Geurts N, Deroost K, Van Aelst I, Verhenne S, Heremans H, Van Damme J, Opdenakker G. Immunopathology and dexamethasone therapy in a new model for malaria-associated acute respiratory distress syndrome. *Am J Respir*

*Crit Care Med* 2010;181:957-968.
49. Rockx B, Baas T, Zornetzer GA, Haagmans B, Sheahan T, Frieman M, Dyer MD, Teal TH, Proll S, van den Brand J, Baric R, Katze MG. Early upregulation of acute respiratory distress syndrome-associated cytokines promotes lethal disease in an aged-mouse model of severe acute respiratory syndrome coronavirus infection. *J Virol* 2009;83:7062-7074.
50. Zhang Y, Sun H, Fan L, Ma Y, Sun Y, Pu J, Yang J, Qiao J, Ma G, Liu J. Acute respiratory distress syndrome induced by a swine 2009 h1n1 variant in mice. *PLoS One* 2012;7:e29347.
51. Fahy RJ, Lichtenberger F, McKeegan CB, Nuovo GJ, Marsh CB, Wewers MD. The acute respiratory distress syndrome: A role for transforming growth factor-beta 1. *Am J Respir Cell Mol Biol* 2003;28:499-503.
52. Daniels CE, Wilkes MC, Edens M, Kottom TJ, Murphy SJ, Limper AH, Leof EB. Imatinib mesylate inhibits the profibrogenic activity of tgf-beta and prevents bleomycin-mediated lung fibrosis. *J Clin Invest* 2004;114:1308-1316.
53. Border WA, Noble NA. Transforming growth factor beta in tissue fibrosis. *N Engl J Med* 1994;331:1286-1292.
54. Kitamura H, Cambier S, Somanath S, Barker T, Minagawa S, Markovics J, Goodsell A, Publicover J, Reichardt L, Jablons D, Wolters P, Hill A, Marks JD, Lou J, Pittet JF, Gauldie J, Baron JL, Nishimura SL. Mouse and human lung fibroblasts regulate dendritic cell trafficking, airway inflammation, and fibrosis through integrin alphavbeta8-mediated activation of tgf-beta. *J Clin Invest* 2011;121:2863-2875.
55. Yang HZ, Wang JP, Mi S, Liu HZ, Cui B, Yan HM, Yan J, Li Z, Liu H, Hua F, Lu W, Hu ZW. Tlr4 activity is required in the resolution of pulmonary inflammation and fibrosis after acute and chronic lung injury. *Am J Pathol* 2012;180:275-292.
56. Hagiwara S, Iwasaka H, Matsumoto S, Noguchi T, Yoshioka H. Association between heat stress protein 70 induction and decreased pulmonary fibrosis in an animal model of acute lung injury. *Lung* 2007;185:287-293.
57. Hilberath JN, Carlo T, Pfeffer MA, Croze RH, Hastrup F, Levy BD. Resolution of toll-like receptor 4-mediated acute lung injury is linked to eicosanoids and suppressor of cytokine signaling 3. *FASEB J* 2011;25:1827-1835.
58. Sakashita A, Nishimura Y, Nishiuma T, Takenaka K, Kobayashi K, Kotani Y, Yokoyama M. Neutrophil elastase inhibitor (sivelestat) attenuates subsequent ventilator-induced lung injury in mice. *Eur J Pharmacol* 2007;571:62-71.
59. D'Alessio FR, Tsushima K, Aggarwal NR, West EE, Willett MH, Britos MF, Pipeling MR, Brower RG, Tuder RM, McDyer JF, King LS. Cd4+cd25+foxp3+ tregs resolve experimental lung injury in mice and are present in humans with acute lung injury. *J Clin Invest* 2009;119:2898-2913.

60. Venet F, Chung CS, Huang X, Lomas-Neira J, Chen Y, Ayala A. Lymphocytes in the development of lung inflammation: A role for regulatory cd4+ t cells in indirect pulmonary lung injury. *J Immunol* 2009;183:3472-3480.
61. Shenkar R, Coulson WF, Abraham E. Anti-transforming growth factor-beta monoclonal antibodies prevent lung injury in hemorrhaged mice. *Am J Respir Cell Mol Biol* 1994;11:351-357.